\documentstyle[amssymb,thmsa,sw20mitp]{article}

\input{tcilatex}
\begin{document}

\title{On the balance between Emigration and Immigration as Random Walks on the
non-negative integers}
\author{Thierry E. Huillet \\
Laboratoire de Physique Th\'{e}orique et Mod\'{e}lisation,\\
CY Cergy Paris Universit\'{e}, CNRS UMR-8089,\\
2 avenue Adolphe-Chauvin, 95302 Cergy-Pontoise, FRANCE\\
Thierry.Huillet@cyu.fr}
\maketitle

\begin{abstract}
Life is on the razor's edge as resulting from competitive birth and death
random forces. We illustrate this aphorism in the context of three Markov
chain population models where systematic random immigration events promoting
growth are simultaneously balanced with random emigration ones provoking
thinning. The origin of mass removals are either determined by external
demands or by aging leading to different conditions of stability.\newline



\end{abstract}

\section{Introduction}

Catastrophic events striking some population can cause major death toll and
they hopefully occur rarely at hectic times. At calm times, the population
can simply grow safely, maybe at a random pace. Catastrophe models are based
on this idea that birth and death are exclusive events. The binomial
catastrophe model is when, on a catastrophic event, the individuals of the
current population each can die or survive{\bf \ }in an independent and even
way with some probability, resulting in a drastic depletion of individuals
at each catastrophic step. For such systems, there is then a competition
between random growth and declining forces, resulting in a subtle balance of
the two. They can be handled in the context of discrete-time Markov chains
on the non-negative integers, see (\cite{Neuts}\cite{Bro}\cite{Ben}\cite{Fon}%
).

In \cite{Neuts}, a catastrophe random walk model was also introduced in
which the origin of the removal of individuals was based on a ``truncated
geometric'' model. In words, if a geometric catastrophic event occurs, given
the population is in some state, its size further shrinks by a random
(geometrically distributed) amount so long as this amount does not exceed
the current state; if it does the population size is set to $\left\{
0\right\} $,\ a disastrous event, \cite{Hui}. So, the random sequential
thinning of the population keeps going on but is stopped as soon as the
current population size is exhausted. At calm times, the population is
incremented by a random amount. The geometric effect corresponds to softer
depletion issues than for the binomial model. This random walk may also be
viewed as giving the size of some population facing (possibly) accidentally
a steady random geometric demand of emigrants from outside or alternatively
being revitalized by a steady number of immigrants. This model was recently
further analyzed in \cite{Hu}. More examples and motivations can be found in 
\cite{Art}\cite{CP}, essentially in continuous-times.

Putting aside the catastrophe idea, one can think of a similar random walk
process now giving the size of some population facing systematically a
steady random demand of emigrants from outside and simultaneously being
revitalized by a steady number of immigrants. Alternatively, this Markov
chain is a model for the state of some stock resulting from the competition
between successive random supplies systematically balanced by simultaneous
truncated geometric random demands. This results in a Markov chain of a new
type.

All such Markov chains models have been designed in an attempt to explain
the transient and large-time behavior of the populations size. Some results
concern the evaluation of the risk of extinction and the distribution of the
population size in the case of total disasters where all individuals in the
population are removed simultaneously, (\cite{Sw}\cite{Hui}). Such Markov
chains are random walks on the non-negative integers (as a semigroup) which
differ from standard random walks on the integers (as a group) in that a
one-step move down from some positive integer cannot take the walker to a
negative state, resulting in transition probabilities being state-dependent.

In Section $2$ of this work, we first study the variation of the Neuts'
truncated geometric catastrophe model in discrete-time, in which input and
output can occur simultaneously. It is also a Markov chain on the
non-negative integers.

Using a generating function approach, we first discuss the condition under
which this process is recurrent (either positive or null) or transient. The
recurrence/transience phase transition is sharp and it easily occurs, due to
moderate depletion at shrinkage steps of this model.

To this end, a recurrence relation for the probability generating function
of this process is first derived (Proposition 1). It is used to discriminate
the phase transition between recurrent and transient regimes.

- In the positive-recurrent (subcritical) regime, we describe the shape and
features of the invariant probability measure (Theorem 2). We emphasize that
in the null-recurrent (critical) regime, no non trivial invariant measure
exists. We then show how to compute the double (space/time) generating
functional of the process (Proposition 3). Using this representation, we
derive the generating functions of the first return time to zero (the length
of the excursions) and of the first {\em local} extinction time (else first
hitting time of $0$) when the process is started at $x_{0}>0$. In the
recurrent regime, a first local extinction occurs with probability $1$. The
analyses rely on the existence of a `key' function which is the inverse of
the singularity of the double generating function of the process. The
Lagrange inversion formula yields a power-series expansion of the `key'
function (Proposition 4). It is used to prove a decomposition result for the
first local extinction time (Theorem 5).

- In the transient (supercritical) regime, after a finite number of visits
to $\left\{ 0\right\} $, the chain drifts to $\infty $. We first emphasize
that no non trivial invariant measure exists either. Using the expression of
the double generating functional of the process, we get access to a precise
large deviation result (Proposition 6). The harmonic (or scale) function of
a version of the process forced to be absorbed in state $\left\{ 0\right\} $
is then used to give an expression of the probability that extinction occurs
before explosion (Proposition 7). \newline

In Section $3$, two related emigration/immigration models are analyzed.

One is a model related to the truncated geometric emigration/immigration one
of Section $2$; it was recently introduced in \cite{BS}. In one of its
interpretations, the state variable is now the number of individuals that
could potentially be released in the latter truncated geometric model: the
chain's state is the system's capacity of release. This `dual' chain is
analyzed along similar lines as before and it is shown to exhibit very
different statistical features; in particular it is emphasized that the
corresponding Markov chain is always positive-recurrent (subcritical)
whatever the law of the input (Proposition 8); a remarkable stability
property. Using the expression of its double generating function
(Proposition 9), we further show that, in this context, the first hitting
time of $0$ when starting from $x_{0}\geq 1$, is independent of $x_{0}$ and
geometrically distributed (Corollary 10).

For comparison, we also briefly revisit the subcritical
Bienaym\'{e}-Galton-Watson branching process with immigration for which the
origin of the pruning of individuals is rather internal and due to the
intrinsic unbalance of birth and death events inside the population. In such
models, both emigration and immigration also can occur simultaneously as
well and so they have not the catastrophe flavor either. The conditions for
the recurrence/transience transition are recalled and detailed; they are
weak due to massive depletion of individuals at shrinkage steps. In contrast
with the truncated geometric model, the critical regime for the underlying
branching process exhibits a non-trivial asymptotic behavior, following
Seneta's results, \cite{Seneta}.

All models involve two stationary stochastic sources, one determining the
thinning of the population, the other competing one, additive, its growth.
In this sense, they are bi-stochastic. Semi-stochastic growth/collapse or
decay/surge Piecewise Deterministic Markov Processes in the same vein,
although in the continuum, were recently considered in \cite{GHL1}\cite{GHL2}%
, following the work \cite{EK2006}, where physical applications were
developed, such as queueing processes arising in the Physics of dams or
stress release issues arising in the Physics of earthquakes. Here, growth
was determined by a deterministic differential equation and the collapse
effect was random and state-dependent.

\section{Truncated geometric models}

In \cite{Hu} the following random walk on the non-negative integers was
revisited, following the work of \cite{Neuts}. It is a Markov chain modeling
the competition between successive random supplies $\beta _{n}$ balanced by
truncated geometric random demands $\delta _{n}$. The random walk process
gives the size $X_{n}$ of some population facing accidentally a steady
random demand $\delta _{n}$ of emigrants from outside or {\em alternatively}
being revitalized by a steady number $\beta _{n}$ of immigrants. The birth
and death sequences were as follows:

$\bullet $ Birth (growth, immigration):

The sequence $\left( \beta _{n}\right) _{n\geq 1}$ was an independent and
identically distributed (i.i.d.) sequence taking values in ${\Bbb N}%
:=\left\{ 1,2,...\right\} ,$ with common law $b_{x}:={\bf P}\left( \beta
=x\right) $, $x\geq 1$, obeying $b_{0}=0$.

$\bullet $ Death (thinning, emigration):

The sequence $\left( \delta _{n}\right) _{n\geq 1}$ was an i.i.d. shifted
geometric$\left( \alpha \right) -$distributed one \footnote{%
A geometric$\left( \alpha \right) $ rv with success probability $\alpha $
takes values in ${\Bbb N}$. A shifted geometric$\left( \alpha \right) $ rv
with success probability $\alpha $ takes values in ${\Bbb N}_{0}$. It is
obtained while shifting the former one by one unit.}, with failure parameter 
$\alpha \in \left( 0,1\right) $, viz. ${\bf P}\left( \delta =x\right)
=\alpha \overline{\alpha }^{x}$, $x\geq 0$ (where $\overline{\alpha }%
:=1-\alpha $). Due to the memory-less property of $\delta $, the parameter $%
\alpha $ is alternatively the (constant) discrete failure rate of $\delta ,$
namely $\alpha ={\bf P}\left( \delta =x\right) /{\bf P}\left( \delta \geq
x\right) $, $x\geq 0.$

The Markov chain under study was given by, \cite{Neuts}{\bf :} 
\begin{equation}
X_{n+1}=\left\{ 
\begin{array}{c}
X_{n}+\beta _{n+1}\text{ with probability }p \\ 
\left( X_{n}-\delta _{n+1}\right) _{+}\text{ with probability }q=1-p,
\end{array}
\right.  \label{G1}
\end{equation}
with alternating{\em ,} so exclusive, immigration and emigration events,
respectively. The sequence $\beta _{n}$ was assumed to be independent of $%
\delta _{n}$ and $X_{n-k}$ for all $k\geq 1.$ The truncated emigration
component in (\ref{G1}) is a non-linear random function of the state $X_{n}$.

At each step $n$, the walker either moves up with probability $p$, the
amplitude of the upward move beings $\beta _{n+1}$ or, given the chain is in
state $x$, the number of step-wise removed individuals is $\delta
_{n+1}\wedge x:=\min \left( \delta _{n+1},x\right) ,$ with probability $q$.

Note that if $X_{n}=0$, then $X_{n+1}=\beta _{n+1}$ with probability $p$
(reflection at $0$) and $X_{n+1}=0$ with probability $q$ (absorption at $0$).%
\newline

{\em Remark:} Considerable simplifications are expected when $\beta =1$ with
probability $1$: in this case, the transition matrix $P$ associated to (\ref
{G1}) is of Hessenberg type, that is lower triangular with a non-zero upper
diagonal. The corresponding process is a skip-free to the right Markov
chain. If in addition, $\delta =1$ with probability $1$ (which is ruled out
here because $\delta $ was assumed here geometrically distributed), we are
left with the standard birth and death chain with tridiagonal transition
matrix for which a non-trivial extension would be to have the probabilities $%
p$ and $q$ to move up and down depend on the current state $x,$ \cite{Hu2}.
Conditions of ergodicity of such chains are well-known and have been studied
for long, see \cite{Kar} for example. The Neuts' random walk is thus a
generalized birth and death one with state-dependent probability
transitions. Exclusivity of birth and death events suggest that in a typical
regime, births are at stake, whereas, possibly exceptionally if $q$ is
small, a catastrophic event with death toll occurs, \cite{Hu}, \cite{Hu2}.
In the following version of this model, there is no longer such an
interpretation of rare catastrophic events preventing growth. $\rhd $

\subsection{Simultaneous growth and depletion: a new truncated geometric
model}

We proceed with a similar study of the following modified version of the
Neuts' process, deserving a special study and with specific statistical
features. Consider the time-homogeneous Markov chain now with temporal
evolution{\bf :} 
\begin{equation}
X_{n+1}=\left( X_{n}-\delta _{n+1}\right) _{+}+\beta _{n+1};\text{ }%
X_{0}\geq 0\text{, else}  \label{chain2}
\end{equation}
\[
\Delta X_{n}=X_{n+1}-X_{n}=-X_{n}\wedge \delta _{n+1}+\beta _{n+1} 
\]
One of its possible observable is 
\[
Y_{n+1}=-X_{n}\wedge \delta _{n+1}, 
\]
the amount of individuals that currently effectively moves out the system.
Based on observed $\left( y_{n}\right) $, de-trending can be used to
estimate $\left( x_{n}\right) .$

In contrast with the previous model (\ref{G1}), competing depletion and
growth mechanisms can occur simultaneously; the two are no longer exclusive.
As before, $X_{n}$ may represent the amount of some resource available at
time $n$ or the state of the fortune of some investor facing recurrent
expenses but sustained by recurrent income. Due to the demand $\delta _{n+1}$
on day $n+1$, the stock shrinks by the random amount $S\left( X_{n}\right)
:=X_{n}\wedge \delta _{n+1}$ and, {\em concomitantly}, there is a
simultaneous production $\beta _{n+1}$ of this resource, used to face
forthcoming demands. One can also loosely think of $X_{n}$ as a model of the
height of a random polymer in a stationary random environment, (see \cite
{Bha} and References therein). Polymer models as `classical'
space-inhomogeneous birth and death Markov chains on ${\Bbb N}_{0}:=\left\{
0,1,2,...\right\} $ (with moves up and down only by one unit) were
considered in \cite{AK} for example. They exhibit a pinning/depinning
transition.\newline

The birth and death sequences of our model are now chosen as follows:

$\bullet $ Growth [input]:

$\left( \beta _{n}\right) _{n\geq 1}$ is an i.i.d. sequence now taking
values in ${\Bbb N}_{0},$ with common law $b_{x}:={\bf P}\left( \beta
=x\right) $, $x\geq 0$. We assume $0<b_{0}<1.$ We shall let $\phi _{\beta
}\left( z\right) :={\bf E}\left( z^{\beta }\right) $ be the common
probability generating function (p.g.f.) of the $\beta $'s$.$ There is now a
positive probability $b_{0}$ that $\beta =0.$

$\bullet $ Depletion [output] (thinning):

$\left( \delta _{n}\right) _{n\geq 1}$ is an i.i.d. shifted geometric$\left(
\alpha \right) -$distributed sequence (independent of $\left( \beta
_{n}\right) _{n\geq 1}$), with failure parameter $\alpha \in \left(
0,1\right) $. Each $\delta $ then takes values in ${\Bbb N}_{0}$ with common
law ${\bf P}\left( \delta =x\right) =:d_{x}=\alpha \overline{\alpha }^{x}$, $%
x\geq 0$ and mean $\mu _{\delta }:={\bf E}\left( \delta \right) =\overline{%
\alpha }/\alpha $. State $0$ is not absorbing (else reflecting) because $%
{\bf P}\left( X_{1}=0\mid X_{0}=0\right) =b_{0}<1:$ after some geometric
random number of steps, the walker is bounced back inside the state-space$.$
There is a positive probability $d_{0}=\alpha $ that $\delta =0$.

The sequence $\beta _{n}$ is assumed to be independent of $\delta _{n}$ and $%
X_{n-k}$ for all $k\geq 1.$

The one-step stochastic transition matrix $P$ (obeying $P{\bf 1}={\bf 1}$,
where ${\bf 1}$ is a column vector of ones) of the Markov chain $X_{n}$ is
(with $d_{z}=0$ if $z<0$ and the convention $\sum_{z=a}^{b}=0$ if $b<a$): 
\begin{equation}
P\left( x,y\right) =\sum_{z=x-y}^{x-1}b_{y-x+z}d_{z}+{\bf P}\left( \delta
\geq x\right) b_{y}\text{, }x,y\geq 0,  \label{GG2}
\end{equation}
Note $P\left( 0,y\right) =b_{y}$, $y\geq 0$ and $P\left( x,0\right) ={\bf P}%
\left( \delta \geq x\right) b_{0}$. In particular also (if emigration equals
immigration) $P\left( x,x\right) =\sum_{z=0}^{x-1}b_{z}d_{z}+{\bf P}\left(
\delta \geq x\right) b_{x}$ are its diagonal matrix elements$.$ This Markov
chain is time-homogeneous, irreducible and aperiodic if and only if $%
0<b_{0}<1$, with state $\left\{ 0\right\} $ reflecting. As a result, all
states $x=\left\{ 0,1,2,...\right\} $ are either recurrent or transient.
Note that the likelihood of a path $\left( x_{0},...,x_{n}\right) $%
\[
\prod_{m=1}^{n}P\left( x_{m-1},x_{m}\right) 
\]
has a very complex structure.\newline

{\em Remark:} (\ref{chain2}) suggests the continuous-time version of this
counting process 
\[
dX_{t}=-X_{t_{-}}\wedge \delta _{t}\cdot dt+dN_{t} 
\]
where $\delta _{t}\stackrel{d}{\sim }\delta $ for all $t\geq 0$ is a
strongly stationary geometric process and $N_{t}$ a compound-Poisson process
with discrete jumps' amplitudes $\beta .$ We shall not run here into the
analysis of this process, postponing it to future work.

A continuous space-time version of this process with state-space $\left[
0,\infty \right) $ would be when $\delta _{t}\stackrel{d}{\sim }\delta $ is
a strongly stationary process with $\delta $ exponentially distributed and
when allowing $\beta $ now to take values in the half-line or, more
generally, when allowing $N_{t}$ to be a general L\'{e}vy subordinator. $%
\rhd $

\subsection{The recurrence relation for the p.g.f. of $X_{n}$}

For any given $X_{n}=x,$ the shrinkage random variable (r.v.): $S\left(
x\right) :=\left( x-\delta \right) _{+}$ obeys $S\left( x\right) \leq x$,
having support on $\left\{ 0,...,x\right\} $. Hence $S\left( X_{n}\right)
\leq X_{n}$ almost surely (a.s.). With $\varepsilon _{x}$ the Dirac mass at $%
x$, the law of $S\left( x\right) $ is obtained as follows: 
\begin{eqnarray*}
&&S\left( x\right) \stackrel{d}{\sim }\overline{\alpha }^{x}\varepsilon
_{0}+\sum_{y=0}^{x-1}\alpha \overline{\alpha }^{y}\varepsilon _{x-y} \\
{\bf E}\left( z^{S\left( x\right) }\right) &=&\overline{\alpha }^{x}+\alpha
\sum_{y=0}^{x-1}\overline{\alpha }^{y}z^{x-y}=\overline{\alpha }^{x}+\alpha 
\overline{\alpha }^{x}z\frac{1-\left( z/\overline{\alpha }\right) ^{x}}{%
\overline{\alpha }-z}
\end{eqnarray*}
leading to the local drift and variance terms 
\begin{eqnarray*}
f\left( x\right) &:&={\bf E}\left( X_{n+1}-x\mid X_{n}=x\right) ={\bf E}%
\left( S\left( x\right) \right) -x+{\bf E}\left( \beta \right) =-\frac{%
\overline{\alpha }}{\alpha }\left( 1-\overline{\alpha }^{x}\right) +{\bf E}%
\left( \beta \right) \\
&\sim &-\frac{\overline{\alpha }}{\alpha }+{\bf E}\left( \beta \right) \text{
as }x\rightarrow \infty . \\
\sigma ^{2}\left( x\right) &:&=\sigma ^{2}\left( \left( X_{n+1}-x\right)
\mid X_{n}=x\right) =\sigma ^{2}\left( S\left( x\right) \right) +\sigma
^{2}\left( \beta \right) \\
&=&\frac{\overline{\alpha }}{\alpha ^{2}}-\frac{2x+1}{\alpha }\overline{%
\alpha }^{x+1}-\left( \frac{\overline{\alpha }^{x+1}}{\alpha }\right)
^{2}+\sigma ^{2}\left( \beta \right) \\
&\sim &\sigma ^{2}\left( \delta \right) +\sigma ^{2}\left( \beta \right) 
\text{ as }x\rightarrow \infty .
\end{eqnarray*}
If $\mu _{\beta }:={\bf E}\left( \beta \right) <\infty $ and $x$ is large,
the walker `feels' the constant drift $f\left( x\right) \stackunder{%
x\rightarrow \infty }{\sim }-\frac{\overline{\alpha }}{\alpha }+{\bf E}%
\left( \beta \right) =\mu _{\beta }-\mu _{\delta }$, positive or negative,
depending on $\mu _{\beta }\gtrless \mu _{\delta }.$ Fluctuations are of
order $\sigma ^{2}\left( \delta \right) +\sigma ^{2}\left( \beta \right) $
for large $x$. Averaging over $X_{n}$, we obtain

\begin{proposition}
\begin{equation}
\Phi _{n+1}\left( z\right) :={\bf E}\left( z^{X_{n+1}}\right) =\phi _{\beta
}\left( z\right) \sum_{x\geq 0}{\bf P}\left( X_{n}=x\right) \left( \overline{%
\alpha }^{x}+\alpha \overline{\alpha }^{x}z\frac{1-\left( z/\overline{\alpha 
}\right) ^{x}}{\overline{\alpha }-z}\right)   \label{GG0}
\end{equation}
\begin{eqnarray*}
=\phi _{\beta }\left( z\right) \left( \Phi _{n}\left( \overline{\alpha }%
\right) +\frac{\alpha z}{\overline{\alpha }-z}\left( \Phi _{n}\left( 
\overline{\alpha }\right) -\Phi _{n}\left( z\right) \right) \right)  \\
=\phi _{\beta }\left( z\right) \left( \frac{\overline{\alpha }\left(
1-z\right) }{\overline{\alpha }-z}\Phi _{n}\left( \overline{\alpha }\right) -%
\frac{\alpha z}{\overline{\alpha }-z}\Phi _{n}\left( z\right) \right) 
\end{eqnarray*}
\end{proposition}

With ${\bf \pi }_{n}:=\left( {\bf P}_{x_{0}}\left( X_{n}=0\right) ,{\bf P}%
_{x_{0}}\left( X_{n}=1\right) ,...\right) ^{\prime }$ the column vector%
\footnote{%
In the sequel, a boldface variable, say ${\bf x}$, will represent a
column-vector so that its transpose, say ${\bf x}^{\prime }$, will be a
row-vector.} of the states' occupation probabilities at time $n$, ${\bf \pi }%
_{n+1}^{\prime }={\bf \pi }_{n}^{\prime }P$, $X_{0}\stackrel{d}{\sim }\delta
_{x_{0}},$ is the master equation of its temporal evolution and (\ref{GG0})
is the corresponding temporal evolution of $\Phi _{n}\left( z\right)
=\sum_{x\geq 0}z^{x}{\bf P}_{x_{0}}\left( X_{n}=x\right) .$

\subsection{Positive recurrence and the invariant probability measure
(subcritical regime)}

If $\Phi _{\infty }\left( z\right) $ were to exist, it should solve $\Phi
_{\infty }\left( z\right) =\phi _{\beta }\left( z\right) \left( \frac{%
\overline{\alpha }\left( 1-z\right) }{\overline{\alpha }-z}\Phi _{\infty
}\left( \overline{\alpha }\right) -\frac{\alpha z}{\overline{\alpha }-z}\Phi
_{\infty }\left( z\right) \right) $ which is 
\begin{equation}
\Phi _{\infty }\left( z\right) =\frac{\overline{\alpha }\Phi _{\infty
}\left( \overline{\alpha }\right) \phi _{\beta }\left( z\right) \left(
1-z\right) }{1-z-\alpha \left( 1-z\phi _{\beta }\left( z\right) \right) }=%
\frac{{\cal N}\left( z\right) }{{\cal D}\left( z\right) }.  \label{GG0b}
\end{equation}
Note that $\Phi _{\infty }\left( 0\right) =b_{0}\Phi _{\infty }\left( 
\overline{\alpha }\right) $ would be the probability that the chain is
asymptotically in state $\left\{ 0\right\} .$ The solution $\Phi _{\infty
}\left( 0\right) =\Phi _{\infty }\left( \overline{\alpha }\right) =0$
(suggesting transience) is incompatible with recurrence and the fact that
state $\left\{ 0\right\} $ is non absorbing$.$ The chain (\ref{chain2})
being irreducible and aperiodic, it is sufficient for its
positive-recurrence to show that $\Phi _{n}\left( 0\right) $ approaches a
positive limit as $n\rightarrow \infty $, since $\Phi _{n}\left( 0\right) =%
{\bf P}\left( X_{n}=0\right) =P^{n}\left( x_{0},0\right) $ is the $n$-step
transition probability from state $x_{0}$ to state $\left\{ 0\right\} $ of
the chain.

Suppose $X_{\infty }\stackrel{d}{\sim }C\delta _{0}+\left( 1-C\right) \delta
_{\infty }$ for some $C\in \left( 0,1\right) .$ Then $\Phi _{\infty }\left(
z\right) =C$ for all $z\in \left[ 0,1\right] $ and 
\[
1=\frac{\phi _{\beta }\left( z\right) \left( 1-z\right) }{1-z-\alpha \left(
1-\phi _{\beta }\left( z\right) \right) } 
\]
which cannot be true unless $\phi _{\beta }\left( z\right) =1$ ($\beta 
\stackrel{d}{\sim }\delta _{0}$), which is ruled out. However, $\Phi
_{\infty }\left( z\right) \equiv 0$ for all $z\in \left[ 0,1\right] $ ($C=0$
and $X_{\infty }\stackrel{d}{\sim }\delta _{\infty }$) is a possible
solution though, corresponding to a transient chain at $\infty $.

Suppose $\phi _{\beta }\left( z\right) =1-\lambda \left( 1-z\right) ^{\theta
}+o\left( \left( \left( 1-z\right) ^{\theta }\right) \right) $\ with $%
\lambda ,\theta \in \left( 0,1\right) $, so that $\mu _{\beta }:=\phi
_{\beta }^{\prime }\left( 1\right) ={\bf E}\left( \beta \right) =\infty .$\
Then, as $z\rightarrow 1$%
\begin{eqnarray*}
{\cal N}\left( z\right) &\sim &\Phi _{\infty }\left( \overline{\alpha }%
\right) \left( 1-z\right) \\
{\cal D}\left( z\right) &\sim &\alpha \lambda \left( 1-z\right) ^{\theta } \\
\Phi _{\infty }\left( z\right) &\sim &\frac{\Phi _{\infty }\left( \overline{%
\alpha }\right) }{\alpha \lambda }\left( 1-z\right) ^{1-\theta }\rightarrow 0
\end{eqnarray*}
The chain asymptotically drifts to $\infty $\ with probability $1$.

Suppose $\mu _{\beta }:=\phi _{\beta }^{\prime }\left( 1\right) ={\bf E}%
\left( \beta \right) <\infty $. In that case, the numerator ${\cal N}$\ and
denominator ${\cal D}$\ both tend to $0$\ as $z\rightarrow 1$\ while their
ratio $\Phi _{\infty }\left( z\right) $\ tends to some constant $C.$\
Indeed, as $z\rightarrow 1$, 
\begin{eqnarray*}
{\cal N}\left( z\right) &\sim &\overline{\alpha }\Phi _{\infty }\left( 
\overline{\alpha }\right) \left( 1-z\right) \left( 1-\mu _{\beta }\left(
1-z\right) \right) \\
{\cal D}\left( z\right) &\sim &\left( 1-z\right) \left( \overline{\alpha }%
-\alpha \mu _{\beta }\right) \\
\Phi _{\infty }\left( z\right) &\sim &C:=\overline{\alpha }\Phi _{\infty
}\left( \overline{\alpha }\right) /\left( \overline{\alpha }-\alpha \mu
_{\beta }\right) =\frac{\Phi _{\infty }\left( \overline{\alpha }\right) }{%
1-\mu _{\beta }/\mu _{\delta }}
\end{eqnarray*}
With $^{\prime }$ denoting derivative with respect to $z$, L'Hospital rule
yields$\frac{{\cal N}^{\prime }\left( z\right) }{{\cal D}^{\prime }\left(
z\right) }\rightarrow \frac{\Phi _{\infty }\left( \overline{\alpha }\right) 
}{1-\mu _{\beta }/\mu _{\delta }}$\ as $z\rightarrow 1$ and does not allow
to fix $\Phi _{\infty }\left( \overline{\alpha }\right) $.

There is however a proper p.g.f. solution to (\ref{GG0b}) if and only if $%
\mu _{\beta }<\mu _{\delta }$ and $\Phi _{\infty }\left( \overline{\alpha }%
\right) =1-\mu _{\beta }/\mu _{\delta }$. We shall call this regime the
subcritical regime. Indeed, 
\begin{equation}
\Phi _{\infty }\left( 0\right) =\pi \left( 0\right) =b_{0}\Phi _{\infty
}\left( \overline{\alpha }\right) =b_{0}\left( 1-\mu _{\beta }/\mu _{\delta
}\right)  \label{GGG0}
\end{equation}
\ is then the well-behaved probability that the chain is asymptotically in
state $\left\{ 0\right\} $, so that, with $\tau _{0,0}$ the first return
time to $0$, by Kac's theorem, \cite{Kac} 
\[
{\bf E}\left( \tau _{0,0}\right) =1/\pi \left( 0\right) <\infty . 
\]
Plugging this particular value of $\Phi _{\infty }\left( \overline{\alpha }%
\right) $ into (\ref{GG0b}), we get 
\[
\Phi _{\infty }\left( z\right) =\Phi _{\infty }\left( \overline{\alpha }%
\right) \phi _{\beta }\left( z\right) \frac{\overline{\alpha }}{1-\alpha 
\frac{1-z\phi _{\beta }\left( z\right) }{1-z}}=:\left( 1-\mu _{\beta }/\mu
_{\delta }\right) \frac{\overline{\alpha }\phi _{\beta }\left( z\right) }{%
1-\alpha \overline{\phi }_{\beta }\left( z\right) } 
\]
where $\overline{\phi }_{\beta }\left( z\right) :=\frac{1-z\phi _{\beta
}\left( z\right) }{1-z}$ is the generating function (g.f.) of the tail
probabilities of the $\beta +1$'s, the shifted $\beta $'s by one unit so
with $\overline{\phi }_{\beta }\left( 1\right) =1+\mu _{\beta }$.

\begin{theorem}
The chain $X_{n}$ (\ref{chain2}) with geometric shrinkage $S$ is ergodic
(positive recurrent) if and only if $\mu _{\beta }<\mu _{\delta }<\infty .$
With $a:=\alpha \overline{\phi }_{\beta }\left( 1\right) \in \left(
0,1\right) $ ($\overline{a}=1-a$), independently of $X_{0}$, the p.g.f. $%
\Phi _{\infty }\left( z\right) $ of $X_{\infty }$ admits the final
expression:{\em \ } 
\begin{equation}
\Phi _{\infty }\left( z\right) =\phi _{\beta }\left( z\right) \frac{1-\alpha 
\overline{\phi }_{\beta }\left( 1\right) }{1-\alpha \overline{\phi }_{\beta
}\left( z\right) }=\phi _{\beta }\left( z\right) \frac{\overline{a}}{%
1-aC\left( z\right) },  \label{GG5}
\end{equation}

where $\frac{\overline{a}}{1-aC\left( z\right) }$ is a compound
shifted-geometric$\left( a\right) $ of the r.v.'s with compounding p.g.f. $%
C\left( z\right) =\overline{\phi }_{\beta }\left( z\right) /\overline{\phi }%
_{\beta }\left( 1\right) .$
\end{theorem}

The compounding p.g.f. $C\left( z\right) $ is the delay p.g.f. of $\beta +1$%
. It has mean $C^{\prime }\left( 1\right) ={\bf E}\left[ \beta \left( \beta
+1\right) \right] /\left[ 2{\bf E}\left( \beta +1\right) \right] <\infty $
if and only if ${\bf E}\left[ \beta ^{2}\right] <\infty .$ From (\ref{GG5}), 
$X_{\infty }$ is the sum of two independent components: one is $\beta $, the
other one is the r.v. with p.g.f. $\overline{a}/\left( 1-aC\left( z\right)
\right) $, so $X_{\infty }$ exceeds $\beta $ and if the r.v. $\beta $ is
infinitely divisible (compound Poisson), so is the r.v. $X_{\infty }$. In
this positive recurrent regime, we have: 
\[
m:={\bf E}\left( X_{\infty }\right) ={\bf E}\left( \beta \right) +\frac{a}{%
\overline{a}}C^{\prime }\left( 1\right) =\mu _{\beta }+\frac{\mu _{\delta }%
{\bf E}\left[ \beta \left( \beta +1\right) \right] }{2\left( 1-\mu _{\beta
}/\mu _{\delta }\right) }\leq \infty 
\]
Inputs $\beta $ with infinite variance yield an invariant probability
measure with infinite mean. An example of such an input $\beta $ with
infinite variance is the one with p.g.f. 
\[
\phi _{\beta }\left( z\right) =1-\mu _{\beta }\left( 1-z\right) +\frac{%
\lambda }{\theta }\left( 1-z\right) ^{\theta }, 
\]
with $\lambda \in \left( 0,1\right) $ and $\theta \in \left( 1,2\right) .$
In this case, $\overline{\phi }_{\beta }\left( z\right) =1+\mu _{\beta }z-%
\frac{\lambda z}{\theta }\left( 1-z\right) ^{\theta -1}$ and $C\left(
z\right) =\overline{\phi }_{\beta }\left( z\right) /\overline{\phi }_{\beta
}\left( 1\right) $ obeys $C^{\prime }\left( 1\right) =\infty .$

\subsubsection{No invariant measure in the null-recurrent case (critical
regime)}

Null-recurrent or transient random walks on a countable state-space may have
or not a stationary measure, (\cite{Harr}, \cite{Har}). If it has, it may
not be unique. Let $\left[ z^{x}\right] \Phi _{\infty }\left( z\right) $ be
the $z^{x}$-coefficient of $\Phi _{\infty }\left( z\right) $ in its series
expansion. If $\mu _{\beta }=\mu _{\delta }$, the critical chain is
null-recurrent. Recalling $\Phi _{\infty }\left( z\right) =\frac{\overline{%
\alpha }\Phi _{\infty }\left( \overline{\alpha }\right) \phi _{\beta }\left(
z\right) \left( 1-z\right) }{1-z-\alpha \left( 1-z\phi _{\beta }\left(
z\right) \right) }$ with $\Phi _{\infty }\left( 0\right) =\pi \left(
0\right) =b_{0}\Phi _{\infty }\left( \overline{\alpha }\right) ,$%
\[
\pi \left( x\right) :={\bf P}_{x_{0}}\left( X_{\infty }=x\right) =\left[
z^{x}\right] \Phi _{\infty }\left( z\right) =\pi \left( 0\right) \frac{%
\overline{\alpha }}{b_{0}}\left[ z^{x}\right] \frac{\phi _{\beta }\left(
z\right) \left( 1-z\right) }{1-z-\alpha \left( 1-z\phi _{\beta }\left(
z\right) \right) },\text{ }x\geq 1. 
\]
so that 
\[
\pi \left( 0\right) =b_{0}\left( 1-\mu _{\beta }/\mu _{\delta }\right)
=0\Rightarrow \pi \left( x\right) =0\text{ for all }x\geq 1. 
\]
The chain (\ref{chain2}) has no non trivial ($\neq {\bf 0}$) invariant
positive measure.

\subsection{The generating functional of the truncated geometric model}

With $X_{0}=x_{0}\geq 1$, fixed, defining the double generating function 
\[
\Phi _{x_{0}}\left( u,z\right) =\sum_{n\geq 0}u^{n}\Phi _{n}\left( z\right)
=\sum_{n\geq 0}u^{n}{\bf E}_{x_{0}}\left( z^{X_{n}}\right) , 
\]
from (\ref{GG0}), we get 
\[
\left( \overline{\alpha }-z\right) \Phi _{x_{0}}\left( u,z\right)
=z^{x_{0}}\left( \overline{\alpha }-z\right) +u\left[ \phi _{\beta }\left(
z\right) \left( \overline{\alpha }\left( 1-z\right) \Phi _{x_{0}}\left( u,%
\overline{\alpha }\right) -\alpha z\Phi _{x_{0}}\left( u,z\right) \right)
\right] 
\]
\begin{equation}
\Phi _{x_{0}}\left( u,z\right) =\frac{z^{x_{0}}\left( \overline{\alpha }%
-z\right) +\overline{\alpha }u\left( 1-z\right) \phi _{\beta }\left(
z\right) \Phi _{x_{0}}\left( u,\overline{\alpha }\right) }{\overline{\alpha }%
-z+\alpha uz\phi _{\beta }\left( z\right) }.  \label{GG10}
\end{equation}
We have, 
\begin{equation}
\Phi _{x_{0}}\left( u,0\right) =G_{x_{0},0}\left( u\right) =ub_{0}\Phi
_{x_{0}}\left( u,\overline{\alpha }\right) ,  \label{GG9a}
\end{equation}
\[
\Phi _{x_{0}}\left( u,\overline{\alpha }\right) =\Phi _{x_{0}}\left(
u,0\right) /\left[ ub_{0}\right] . 
\]
So far, $\Phi _{x_{0}}\left( u,z\right) $ is unknown since it requires the
knowledge of $\Phi _{x_{0}}\left( u,\overline{\alpha }\right) $ or of $\Phi
_{x_{0}}\left( u,0\right) $. $\Phi _{x_{0}}\left( u,z\right) $ has a
singularity at 
\begin{equation}
u\left( z\right) =\left( z-\overline{\alpha }\right) /\left[ \alpha z\phi
_{\beta }\left( z\right) \right] .  \label{GG8a}
\end{equation}
In the recurrent regime ($\mu _{\beta }\leq \mu _{\delta }$), with $%
u^{\prime }\left( \overline{\alpha }\right) =1/\left( \alpha \overline{%
\alpha }\phi _{\beta }\left( \overline{\alpha }\right) \right) >1$, $u\left(
z\right) $ is concave and monotone increasing on the interval $\left[ 
\overline{\alpha },1\right] ,$ with $u^{\prime }\left( 1\right) =\mu
_{\delta }-\mu _{\beta }\geq 0$ (This can be checked from Proposition $4$
stating that its inverse is absolutely monotone as a p.g.f., in particular
increasing and convex). The corresponding range of $u\left( z\right) $ is $%
\left[ 0,1\right] $. The function $u\left( z\right) $ has thus a
well-defined inverse $z\left( u\right) $ which maps $\left[ 0,1\right] $ to $%
\left[ \overline{\alpha },1\right] $; this inverse is monotone increasing
and convex on this interval. We call it the `key' function, due to its
fundamental interest in the sequel.

When in the recurrent regime, $\left[ u^{n}\right] \Phi _{x_{0}}\left(
u,z\right) \rightarrow \Phi _{\infty }\left( z\right) $ as $n\rightarrow
\infty $, $\Phi _{x_{0}}\left( u,z\right) $ should converge as $z\searrow
z\left( u\right) $ for all $u\in \left[ 0,1\right) .$\ So both the numerator 
${\cal N}$ and the denominator ${\cal D}$ of $\Phi _{x_{0}}\left( u,z\right) 
$\ must tend to $0$ as{\em \ }$z\searrow z\left( u\right) $, meaning (by
L'Hospital rule) that 
\[
\lim_{z\searrow z\left( u\right) }\frac{{\cal N}\left( u,z\right) }{{\cal D}%
\left( u,z\right) }=\lim_{z\searrow z\left( u\right) }\frac{{\cal N}^{\prime
}\left( u,z\right) }{{\cal D}^{\prime }\left( u,z\right) }.
\]
Near $z=z\left( u\right) ,$%
\begin{eqnarray*}
{\cal N}\left( u,z\right)  &=&z^{x_{0}}\left( \overline{\alpha }-z\right) +%
\overline{\alpha }\left( 1-z\right) u\phi _{\beta }\left( z\right) \Phi
_{x_{0}}\left( u,\overline{\alpha }\right) =:a\left( z\right) +b\left(
z\right) \Phi _{x_{0}}\left( u,\overline{\alpha }\right)  \\
&\sim &a\left( z\left( u\right) \right) +b\left( z\left( u\right) \right)
\Phi _{x_{0}}\left( u,\overline{\alpha }\right) +\left( z-z\left( u\right)
\right) \left[ a^{\prime }\left( z\left( u\right) \right) +b^{\prime }\left(
z\left( u\right) \right) \Phi _{x_{0}}\left( u,\overline{\alpha }\right)
\right]  \\
{\cal D}\left( u,z\right)  &=&:c\left( z\right) -d\left( z\right) u\sim
\left( z-z\left( u\right) \right) \left[ c^{\prime }\left( z\left( u\right)
\right) -d^{\prime }\left( z\left( u\right) \right) u\right] ,
\end{eqnarray*}
imposing $a\left( z\left( u\right) \right) +b\left( z\left( u\right) \right)
\Phi _{x_{0}}\left( u,\overline{\alpha }\right) =0$. Therefore, 
\[
\Phi _{x_{0}}\left( u,\overline{\alpha }\right) =\frac{z\left( u\right)
^{x_{0}}\left( z\left( u\right) -\overline{\alpha }\right) }{\overline{%
\alpha }u\left( 1-z\left( u\right) \right) \phi _{\beta }\left( z\left(
u\right) \right) }
\]
and, from (\ref{GG9}), 
\begin{equation}
\Phi _{x_{0}}\left( u,0\right) =b_{0}\frac{z\left( u\right) ^{x_{0}}\left(
z\left( u\right) -\overline{\alpha }\right) }{\overline{\alpha }\left(
1-z\left( u\right) \right) \phi _{\beta }\left( z\left( u\right) \right) }%
=:G_{x_{0},0}\left( u\right) \geq 0,  \label{GG8}
\end{equation}
is the Green kernel of this model, for $u\in \left[ 0,1\right] ;$ with $%
x_{0}\geq 1,$ ${\bf P}_{x_{0}}\left( X_{0}=0\right) =0$ and 
\[
G_{x_{0},0}\left( u\right) =\sum_{n\geq 1}u^{n}{\bf P}_{x_{0}}\left(
X_{n}=0\right) =\sum_{n\geq 1}u^{n}P^{n}\left( x_{0},0\right) =\left(
I-uP\right) ^{-1}\left( x_{0},0\right) ,
\]
the Green kernel (resolvent) of the chain at the endpoints $\left(
x_{0},0\right) ,$ \cite{Neveu}$.$ The matrix element $P^{n}\left(
x_{0},0\right) $ is the contact probability at $0$ at time $n$, starting
from $x_{0}\geq 1$. Note from (\ref{GG8}) that $G_{x_{0},0}\left( 1\right)
=\infty $ in the recurrent regime, translating that $X_{n}$ visits state $0$
infinitely often.

From (\ref{GG8}) and (\ref{GG10}), we thus get a closed form expression of $%
\Phi _{x_{0}}\left( u,z\right) $ when $x_{0}\geq 1,$ as:

\begin{proposition}
In the recurrent regime ($\mu _{\beta }\leq \mu _{\delta }$), with $z\left(
u\right) $ the inverse of $u\left( z\right) $ defined in (\ref{GG8a}) and
explicit in (\ref{GGLag}) below, for $u\in \left[ 0,1\right) $ and $z\geq
z\left( u\right) ,$%
\begin{equation}
\Phi _{x_{0}}\left( u,z\right) =\frac{z^{x_{0}}\left( \overline{\alpha }%
-z\right) \left( 1-z\left( u\right) \right) \phi _{\beta }\left( z\left(
u\right) \right) -z\left( u\right) ^{x_{0}}\left( \overline{\alpha }-z\left(
u\right) \right) \left( 1-z\right) \phi _{\beta }\left( z\right) }{\left(
1-z\left( u\right) \right) \left( \overline{\alpha }-z+\alpha uz\phi _{\beta
}\left( z\right) \right) \phi _{\beta }\left( z\left( u\right) \right) }.
\label{GG}
\end{equation}
\end{proposition}

{\em Remark:} If $z=1,$ $\Phi _{x_{0}}\left( u,1\right) =1/\left( 1-u\right) 
$ and $\Phi _{x_{0}}\left( 0,z\right) =z^{x_{0}}$, as required$.$ The
generating function $\Phi _{x_{0}}\left( u,z\right) $\ in (\ref{GG}) is
well-defined when $z>z\left( u\right) $ and possibly when $z=z\left(
u\right) $ as in the recurrent case. $\triangleright $

\subsection{First return time to $0$ (recurrent regime)}

When $x_{0}=0$, observing ${\bf P}_{0}\left( X_{0}=0\right) =1,$%
\begin{equation}
G_{0,0}\left( u\right) =1+\Phi _{0}\left( u,0\right) =1+\frac{b_{0}\left(
z\left( u\right) -\overline{\alpha }\right) }{\overline{\alpha }\left(
1-z\left( u\right) \right) \phi _{\beta }\left( z\left( u\right) \right) }%
\text{, with }G_{0,0}\left( 1\right) =\infty \text{ (}z\left( 1\right) =1%
\text{).}  \label{GG11a}
\end{equation}
This function is the Green kernel at the endpoints $\left( 0,0\right) .$ If $%
n\geq 1,$\ from the recurrence 
\[
{\bf P}_{0}\left( X_{n}=0\right) =:P^{n}\left( 0,0\right) =\sum_{m=1}^{n}%
{\bf P}\left( \tau _{0,0}=m\right) P^{n-m}\left( 0,0\right) {\it ,} 
\]
\ we see, from taking the generating function of both sides and observing
the right hand-side is a convolution, that the p.g.f. $\phi _{0,0}\left(
u\right) ={\bf E}\left( u^{\tau _{0,0}}\right) $\ of the first return time
to $0$, $\tau _{0,0}$\ and $G_{0,0}\left( u\right) $\ are related by the
Feller relation (see \cite{Feller} and \cite{Bing} pp $3-4$\ for example)%
{\bf : }$G_{0,0}\left( u\right) =1+${\bf \ }$G_{0,0}\left( u\right) \phi
_{0,0}\left( u\right) ${\bf . }Hence,{\it \ }with $\phi _{0,0}\left(
0\right) =0$ and $\phi _{0,0}\left( 1\right) =1,$

\begin{equation}
\phi _{0,0}\left( u\right) ={\bf E}\left( u^{\tau _{0,0}}\right) =\frac{%
G_{0,0}\left( u\right) -1}{G_{0,0}\left( u\right) }=\frac{b_{0}\left(
z\left( u\right) -\overline{\alpha }\right) }{b_{0}\left( z\left( u\right) -%
\overline{\alpha }\right) +\overline{\alpha }\left( 1-z\left( u\right)
\right) \phi _{\beta }\left( z\left( u\right) \right) }\text{.}  \label{GG12}
\end{equation}
In particular, observing $z^{\prime }\left( 1\right) =1/u^{\prime }\left(
1\right) =\frac{1}{\mu _{\delta }-\mu _{\beta }}$, we get 
\begin{eqnarray*}
{\bf E}\left( \tau _{0,0}\right) &=&\overline{\alpha }z^{\prime }\left(
1\right) /\left( b_{0}\alpha \right) =\frac{\mu _{\delta }}{b_{0}\left( \mu
_{\delta }-\mu _{\beta }\right) }\text{ if }\mu _{\beta }<\mu _{\delta }%
\text{ (positive recurrence),} \\
&=&\infty \text{ if }\mu _{\delta }=\mu _{\beta }\text{ (null recurrence).}
\end{eqnarray*}
As required from Kac's theorem: ${\bf E}\left( \tau _{0,0}\right) =1/\pi
\left( 0\right) =1/\left[ b_{0}\left( 1-\mu _{\beta }/\mu _{\delta }\right)
\right] ,$ consistently with (\ref{GGG0}).

\subsection{Contact probability at $0$ and first local time to extinction
(recurrent regime)}

With $x_{0}\geq 1$ ($\phi _{x_{0},0}\left( 0\right) =0$), using (\ref{GG8})
and (\ref{GG11a}), 
\begin{equation}
{\bf E}\left( u^{\tau _{x_{0},0}}\right) =\phi _{x_{0},0}\left( u\right) =%
\frac{G_{x_{0},0}\left( u\right) }{G_{0,0}\left( u\right) }  \label{GG12a}
\end{equation}
\[
=\frac{b_{0}z\left( u\right) ^{x_{0}}\left( z\left( u\right) -\overline{%
\alpha }\right) }{b_{0}\left( z\left( u\right) -\overline{\alpha }\right) +%
\overline{\alpha }\left( 1-z\left( u\right) \right) \phi _{\beta }\left(
z\left( u\right) \right) }=\phi _{0,0}\left( u\right) z\left( u\right)
^{x_{0}}, 
\]
gives the p.g.f. of the first hitting time of $0$, starting from $x_{0}\geq
1 $ (the first local extinction time of the chain). We also get

\begin{eqnarray*}
{\bf E}\left( \tau _{x_{0},0}\right) &=&{\bf E}\left( \tau _{0,0}\right)
+x_{0}z^{\prime }\left( 1\right) =\frac{b_{0}x_{0}+\mu _{\delta }}{%
b_{0}\left( \mu _{\delta }-\mu _{\beta }\right) }\text{ if }\mu _{\beta
}<\mu _{\delta }\text{ (positive recurrence),} \\
&=&\infty \text{ if }\mu _{\beta }=\mu _{\delta }\text{ (null recurrence).}
\end{eqnarray*}
If $z^{^{\prime \prime }}\left( 1\right) <\infty $ ($\sigma ^{2}\left( \beta
\right) <\infty $ or $\sigma ^{2}\left( \tau _{0,0}\right) <\infty $), the
variance of $\tau _{x_{0},0}$ in the positive-recurrent regime is finite
whereas if $\sigma ^{2}\left( \beta \right) =\infty $, $\sigma ^{2}\left(
\tau _{x_{0},0}\right) =\infty .$

We observe from (\ref{GG8a}) and (\ref{GG12})

\begin{equation}
\phi _{0,0}\left( u\right) ={\bf E}\left( u^{\tau _{0,0}}\right) =\frac{%
\alpha b_{0}uz\left( u\right) }{\alpha b_{0}uz\left( u\right) +\overline{%
\alpha }\left( 1-z\left( u\right) \right) }\text{,}  \label{GG123}
\end{equation}
so that the laws of $\tau _{x_{0},0}$ and $\tau _{0,0}$ are entirely
determined by the key function $z\left( u\right) .$

The function $u\left( z\right) $ is explicitly defined in (\ref{GG8a}) and $%
z\left( u\right) $ can be obtained from the Lagrange inversion formula as
follows: $u\left( z\right) $ has a power series expansion in the scaled
variable $\widehat{z}:=\left( z-\overline{\alpha }\right) /\alpha $ and 
\[
u\left( \widehat{z}\right) =\frac{\widehat{z}}{\left( \overline{\alpha }%
+\alpha \widehat{z}\right) \phi _{\beta }\left( \overline{\alpha }+\alpha 
\widehat{z}\right) }=:\frac{\widehat{z}}{\widehat{\phi }_{\beta }\left( 
\widehat{z}\right) } 
\]
maps $\left[ 0,1\right] $ to $\left[ 0,1\right] $, with $u\left( \widehat{z}%
=0\right) =0$. By Lagrange inversion formula (\cite{Com}, p. $159$), we can
get $\widehat{z}\left( u\right) $ satisfying $u\left( \widehat{z}\left(
u\right) \right) =u$ as 
\[
\widehat{z}\left( u\right) =\sum_{n\geq 1}u^{n}\left[ u^{n}\right] \widehat{z%
}\left( u\right) \text{ with }\left[ u^{n}\right] \widehat{z}\left( u\right)
=\frac{1}{n}\left[ \widehat{z}^{n-1}\right] \widehat{\phi }_{\beta }\left( 
\widehat{z}\right) ^{n}. 
\]
Finally, with $\widehat{z}\left( 0\right) =0$ and $\widehat{z}\left(
1\right) =1$, $z\left( u\right) =\overline{\alpha }+\alpha \widehat{z}\left(
u\right) $ and we obtain:

\begin{proposition}
It holds that 
\begin{equation}
z\left( u\right) =\overline{\alpha }+\alpha \sum_{n\geq 1}\frac{u^{n}}{n}%
\left[ \widehat{z}^{n-1}\right] \widehat{\phi }_{\beta }\left( \widehat{z}%
\right) ^{n}  \label{GGLag}
\end{equation}
is the series expansion representation of the 'key' function $z\left(
u\right) $. It holds that $\left[ u^{n}\right] z\left( u\right) >0$
translating that $z\left( u\right) $ is a p.g.f..
\end{proposition}

{\em Remark}{\bf : }One can be more precise.{\bf \ }As such indeed,{\bf \ }$%
\widehat{z}\left( u\right) $ can be interpreted as the p.g.f. of the total
progeny of a Bienaym\'{e}-Galton-Watson (BGW) process with branching
mechanism p.g.f. $\widehat{\phi }_{\beta }\left( \widehat{z}\right) =\left( 
\overline{\alpha }+\alpha \widehat{z}\right) \phi _{\beta }\left( \overline{%
\alpha }+\alpha \widehat{z}\right) $, (\cite{Harris}, p. $32$)$;$ $\widehat{%
\phi }_{\beta }\left( \widehat{z}\right) $ is thus the p.g.f. of an
offspring r.v. 
\[
\nu :=\text{ber}\left( \alpha \right) +\sum_{b=0}^{\beta }\text{ber}%
_{b}\left( \alpha \right) ,
\]
where ber$\left( \alpha \right) $ is a Bernoulli$\left( \alpha \right) $
r.v. independent of $\sum_{b=0}^{\beta }$ber$_{b}\left( \alpha \right) $ and
(ber$_{b}\left( \alpha \right) $, $b\geq 0$) are i.i.d. Bernoulli$\left(
\alpha \right) $, independent of $\beta $. Note ${\bf E}\left( \nu \right) =%
\widehat{\phi }_{\beta }^{\prime }\left( 1\right) <1$ (the sub-critical BGW
case) if and only if $\mu _{\beta }<\mu _{\delta }$ in which case $\widehat{z%
}^{\prime }\left( 1\right) =1/\left( 1-{\bf E}\left( \nu \right) \right)
<\infty .$

Finally, $z\left( u\right) $ is a zero-inflated p.g.f. version of $\widehat{z%
}\left( u\right) $. Note also that $z\left( u\right) $ is always a p.g.f.,
regardless of whether the chain (\ref{chain2}) is recurrent or transient.
When the chain is recurrent, the inverse $u\left( z\right) $ of $z\left(
u\right) $ is increasing and concave on its full definition domain $\left[ 
\overline{\alpha },1\right] $ reaching $1$ for the first time at $u=1$,
whereas, as sketched below, in the transient regime, $u\left( z\right) $ is
increasing and concave only on the sub-interval $\left[ \overline{\alpha }%
,z_{*}\right] \subset \left[ \overline{\alpha },1\right] $ with $u^{\prime
}\left( z_{*}\right) =0$ and $u\left( z_{*}\right) >1.$ On the sub-interval $%
\left[ z_{*},1\right] $, $u\left( z\right) $ is decreasing and concave, with 
$u^{\prime }\left( 1\right) =\mu _{\delta }-\mu _{\beta }<0.$ In that case, $%
z\left( u\right) $ is the inverse of the left branch expansion of $u\left(
z\right) $ near $\overline{\alpha }$. $\rhd $\newline

The function $\phi _{0,0}\left( u\right) $ in (\ref{GG123}) is thus a
p.g.f., entirely determined by the p.g.f. $z\left( u\right) $. And $z\left(
u\right) ^{x_{0}}$ is the p.g.f. of the r.v. 
\begin{equation}
\tau _{x_{0}}=\sum_{k=1}^{x_{0}}\tau _{k},  \label{GG12b}
\end{equation}
where $\tau _{k}$ is an i.i.d. sequence with $\tau _{k}\stackrel{d}{\sim }%
\tau $. Note ${\bf E}\left( \tau \right) =\frac{\alpha }{\overline{\alpha }}%
{\bf E}\left( \tau _{0,0}\right) ={\bf E}\left( \tau _{0,0}\right) /\mu
_{\delta }$. We have thus proven

\begin{theorem}
In the recurrent regime, the function $z\left( u\right) $ is the p.g.f. of
some random variable $\tau \geq 0$. With $\tau _{x_{0}}$ defined in (\ref
{GG12b}), the first local time to extinction random variable $\tau _{x_{0},0}
$ is decomposable as 
\begin{equation}
\tau _{x_{0},0}\stackrel{d}{=}\tau _{x_{0}}+\tau _{0,0},  \label{GGG}
\end{equation}
where $\tau _{x_{0}}\geq 0$ and $\tau _{0,0}\geq 1$ are independent. With $%
\phi _{0,0}\left( u\right) $ given by (\ref{GG123}), the p.g.f. of $\tau
_{x_{0},0}$ is given in (\ref{GG12a}) with $z\left( u\right) $ the inverse
(probability generating) function of $u\left( z\right) =\left( z-\overline{%
\alpha }\right) /\left[ \alpha z\phi _{\beta }\left( z\right) \right] $.

In particular, if $\sigma ^{2}\left( \beta \right) <\infty ,$ with $\sigma
^{2}\left( \tau \right) =\frac{\alpha }{\overline{\alpha }}\sigma ^{2}\left(
\tau _{0,0}\right) +\frac{\alpha }{\overline{\alpha }^{2}}{\bf E}\left( \tau
_{0,0}\right) ^{2}<\infty ,$%
\[
\sigma ^{2}\left( \tau _{x_{0},0}\right) =x_{0}\sigma ^{2}\left( \tau
\right) +\sigma ^{2}\left( \tau _{0,0}\right) 
\]
\end{theorem}

{\em Remark}{\bf :} In the recurrent regime, $\tau _{x_{0},0}$ only is the
first {\em local} time to extinction (the first hitting time of $0$); the
chain (\ref{chain2}) being bounced back in the bulk of its definition domain
when it hits $0$, there are infinitely many subsequent local times to
extinction (separating consecutive excursions). Would the chain be forced to
have state $0$ absorbing, this first local time to extinction would become
the ultimate time to extinction. $\rhd $

\subsection{The transient (supercritical) regime}

In the transient regime ($\mu _{\beta }>\mu _{\delta }$), the denominator of 
$\Phi _{x_{0}}\left( u\left( z\right) ,z\right) $\ tends to $0$ as{\em \ }$%
u\rightarrow u\left( z\right) $ and{\em \ }$u\left( z\right) $ does not
cancel the numerator: $u\left( z\right) $ is a true pole of $\Phi
_{x_{0}}\left( u,z\right) $. When $z\in \left[ \overline{\alpha },1\right] $%
, $u\left( z\right) $ is concave but it has a maximum $u\left( z_{*}\right) $
strictly larger than $1$, attained at some $z_{*}$ inside $\left( \overline{%
\alpha },1\right) $, owing to $u\left( \overline{\alpha }\right) =0,$ $%
u\left( 1\right) =1$ and $u^{\prime }\left( 1\right) <0$. We anticipate that
for some constant $C>0,$%
\[
{\bf P}_{x_{0}}\left( X_{n}=0\right) \sim C\cdot u\left( z_{*}\right) ^{-n}%
\text{ as }n\rightarrow \infty , 
\]
stating that $\left\{ X_{n}\right\} $ only visits $0$ a finite number of
times before drifting to $\infty $.

\subsubsection{No invariant measure in the transient regime}

Before turning to this question, let us observe the following: suppose $\mu
_{\beta }>\mu _{\delta }$, so that the super-critical geometric chain is
transient. Would an invariant measure exist, it should satisfy $\pi \left(
0\right) =b_{0}\left( 1-\mu _{\beta }/\mu _{\delta }\right) <0$. The chain
has no non trivial ($\neq {\bf 0}$) invariant positive measure either. It is
not Harris-transient, \cite{Har}.

\subsubsection{Large deviations}

Consider $v\left( z\right) :=1/u\left( z\right) $ where $u\left( z\right) $,
as a pole, cancels the denominator of $\Phi _{x_{0}}\left( u,z\right) $
without cancelling its numerator, so 
\[
v\left( z\right) =\frac{\alpha z\phi _{\beta }\left( z\right) }{z-\overline{%
\alpha }}. 
\]
Over the domain $\overline{\alpha }<z\leq $ $1$, $v\left( z\right) $ is
convex with 
\[
v^{\prime }\left( z\right) =\alpha \frac{\phi _{\beta }^{\prime }\left(
z\right) z\left( z-\overline{\alpha }\right) -\overline{\alpha }\phi _{\beta
}\left( z\right) }{\left( z-\overline{\alpha }\right) ^{2}}\text{ and }%
v^{\prime \prime }\left( z\right) >0, 
\]
because $z\left( u\right) $ being a p.g.f., its inverse $u\left( z\right) $
is concave. We have 
\[
v\left( 1\right) =1\text{ and }v^{\prime }\left( 1\right) =\mu _{\beta }-\mu
_{\delta }, 
\]
and $v^{\prime }\left( 1\right) >0$ if and only if the chain is transient.
In this transient regime, 
\[
\Phi _{n}\left( z\right) ^{1/n}\rightarrow v\left( z\right) \text{ as }%
n\rightarrow \infty . 
\]
Define the logarithmic generating function 
\[
F\left( \lambda \right) :=-\log v\left( e^{-\lambda }\right) =\log u\left(
e^{-\lambda }\right) . 
\]

The function $F\left( \lambda \right) $ is concave on its definition domain $%
\lambda \in \left[ 0,-\log \overline{\alpha }\right) $. Starting from $%
F\left( 0\right) =0$, it is first increasing, attains a maximum and then
decreases to $-\infty $ while crossing zero in between. Therefore, there
exists $\lambda ^{*}\in \left( 0,-\log \overline{\alpha }\right) $ such that 
$x_{*}=F^{\prime }\left( \lambda ^{*}\right) =0$. With $x\in \left(
F^{\prime }\left( -\log \overline{\alpha }\right) ,F^{\prime }\left(
0\right) \right] $, define 
\begin{equation}
f\left( x\right) =\inf_{0\leq \lambda <-\log \overline{\alpha }}\left(
x\lambda -F\left( \lambda \right) \right) \leq 0,  \label{GLeg}
\end{equation}
the Legendre conjugate of $F\left( \lambda \right) $. The variable $x$ is
Legendre conjugate to $\lambda $ with $x=F^{\prime }\left( \lambda \right) $
and $\lambda =f^{\prime }\left( x\right) $. Note $F^{\prime }\left( -\log 
\overline{\alpha }\right) =-\infty $ and $F^{\prime }\left( 0\right)
=v^{\prime }\left( 1\right) =\mu _{\beta }-\mu _{\delta }>0.$ On its
definition domain, $f\left( x\right) \leq 0$ is increasing and concave,
starting from $f\left( -\infty \right) =-\infty $ and ending with $f\left(
F^{\prime }\left( 0\right) \right) =0$ where $f^{\prime }\left( F^{\prime
}\left( 0\right) \right) =0$. From \cite{Cra}\cite{Var}, we get:

\begin{proposition}
For those $x$ in the range $\left[ x_{*}=0,F^{\prime }\left( 0\right)
\right] $ and for any $x_{0}>0,$ 
\begin{equation}
\frac{1}{n}\log {\bf P}_{x_{0}}\left( \frac{1}{n}X_{n}\leq x\right) 
\stackunder{n\rightarrow \infty }{\rightarrow }f\left( x\right) .
\label{GLD}
\end{equation}
\end{proposition}

In particular, at $x=F^{\prime }\left( 0\right) =v^{\prime }\left( 1\right)
>0$ where $f\left( F^{\prime }\left( 0\right) \right) =-F\left( 0\right) =0$%
, we get 
\[
\frac{1}{n}X_{n}\stackrel{a.s.}{\rightarrow }r:=v^{\prime }\left( 1\right) 
\text{ as }n\rightarrow \infty , 
\]
giving the rate $r$ at which $X_{n}$ drifts to $\infty $. To keep $%
x=F^{\prime }\left( \lambda \right) $ in the non-negative range $\left[
x_{*}=0,F^{\prime }\left( 0\right) \right] $, the range of $\lambda $ should
then equivalently be restricted to $\left[ 0,\lambda ^{*}\right] $. We
clearly have $f\left( x_{*}\right) =-F\left( \lambda ^{*}\right) <0$ and,
from (\ref{GLD}), 
\begin{equation}
-\frac{1}{n}\log {\bf P}_{x_{0}}\left( \frac{1}{n}X_{n}\leq 0\right) =-\frac{%
1}{n}\log {\bf P}_{x_{0}}\left( X_{n}=0\right) \rightarrow F\left( \lambda
^{*}\right) \text{ as }n\rightarrow \infty .  \label{GDec}
\end{equation}
This shows the rate at which ${\bf P}_{x_{0}}\left( X_{n}=0\right) $ decays
exponentially with $n$. Equivalently, with $z_{*}=e^{-\lambda ^{*}}$, 
\[
{\bf P}_{x_{0}}\left( X_{n}=0\right) \sim C\cdot u\left( z_{*}\right) ^{-n}, 
\]
as initially guessed.

When $\mu _{\beta }>\mu _{\delta }$, the number $z_{*}$ inside $\left( 
\overline{\alpha },1\right) $ exists and it is characterized by $u^{\prime
}\left( z_{*}\right) =0$ where $u\left( z\right) =\left( z-\overline{\alpha }%
\right) /\left[ \alpha z\phi _{\beta }\left( z\right) \right] ,$ hence by 
\[
\frac{\phi _{\beta }^{\prime }\left( z_{*}\right) }{\phi _{\beta }\left(
z_{*}\right) }=\frac{\overline{\alpha }}{z_{*}\left( z_{*}-\overline{\alpha }%
\right) },\text{ }u\left( z_{*}\right) >1. 
\]

When $\mu _{\beta }>\mu _{\delta }$, the Green series $\Phi _{x_{0}}\left(
1,0\right) =G_{x_{0},0}\left( 1\right) =\sum_{n\geq 1}{\bf P}_{x_{0}}\left(
X_{n}=0\right) $ is summable for all $x_{0}\geq 0$ (translating that $X_{n}$
visits $0$ only a finite number of times). In the transient regime, from (%
\ref{GG12a}), 
\[
{\bf P}\left( \tau _{x_{0},0}<\infty \right) =\phi _{x_{0},0}\left( 1\right)
=\frac{G_{x_{0},0}\left( 1\right) }{G_{0,0}\left( 1\right) }<1. 
\]
\newline

{\em The }$\beta -${\em geometric example:} Suppose $\phi _{\beta }\left(
z\right) =1/\left[ 1+\mu _{\beta }\left( 1-z\right) \right] .$ Then $z_{*}$
is characterized by the quadratic equation 
\[
\frac{\mu _{\beta }}{1+\mu _{\beta }\left( 1-z_{*}\right) }=\frac{\overline{%
\alpha }}{z_{*}\left( z_{*}-\overline{\alpha }\right) }, 
\]
whose positive solution is 
\[
z_{*}=\sqrt{\overline{\alpha }\left( 1+1/\mu _{\beta }\right) }. 
\]
We have $\overline{\alpha }<z_{*}<1$ if and only if $\mu _{\beta }>\mu
_{\delta }$ (the supercritical regime). On the range $\overline{\alpha }<z<1$%
, we have 
\begin{eqnarray*}
u^{\prime }\left( z\right) &=&\frac{\overline{\alpha }\left( 1+\mu _{\beta
}\right) -\mu _{\beta }z^{2}}{\alpha z^{2}} \\
u^{\prime \prime }\left( z\right) &=&-\frac{2\overline{\alpha }\left( 1+\mu
_{\beta }\right) }{\alpha z^{3}}<0\text{,}
\end{eqnarray*}
showing that $u\left( z\right) $ has the announced properties. $\diamond $

\subsubsection{The scale function and the extinction probability (transient
case)}

In the transient regime, the chain $\left\{ X_{n}\right\} $ started at $x>0$
can drift to $\infty $ before it first hits $0$. There is thus only a
probability smaller than $1$ that $\left\{ X_{n}\right\} $ gets extinct for
the first time.

In the transient regime, let then $P\rightarrow P^{\text{abs}}$ with the
modification $P^{\text{abs}}\left( 0,y\right) =\delta _{0,y},$ forcing state 
$\left\{ 0\right\} $ in $P$ to be absorbing, corresponding to: $%
X_{n}\rightarrow Y_{n}:=X_{n\wedge \tau _{x,0}}$. The matrix $P^{\text{abs}}$
is stochastic but non irreducible, having state $\left\{ 0\right\} $ as an
absorbing class. The harmonic (or scale) sequence ${\bf h}$ solves: 
\begin{equation}
P^{\text{abs}}{\bf h}={\bf h},  \label{G17}
\end{equation}
where ${\bf h=}\left( h\left( 0\right) ,h\left( 1\right) ,...\right)
^{\prime }$ is a column vector. With $y\gg 1$ and $x\in \left\{
1,...,y-1\right\} ,$ let 
\begin{eqnarray*}
\tau _{x,y} &=&\inf \left( n\geq 0:X_{n}\geq y{\Bbb \mid }\text{ }%
X_{0}=x\right) \text{ if such an }n\text{ exists,} \\
&=&+\infty \text{ if not.}
\end{eqnarray*}
By induction, with $h\left( 0\right) =1$ and with $\tau _{x}^{\left(
y\right) }:=\tau _{x,0}\wedge \tau _{x,y}$%
\[
\forall x\in \left\{ 1,...,y-1\right\} \text{, }\forall n\geq 0\text{: }%
h\left( x\right) ={\bf E}h\left( Y_{n\wedge \tau _{x,y}}\right) ={\bf E}%
h\left( X_{n\wedge \tau _{x}^{\left( y\right) }}\right) . 
\]
The harmonic function on $\left\{ 1,...,y\right\} $, makes $h\left(
X_{n\wedge \tau _{x}^{\left( y\right) }}\right) $ a martingale, \cite{Nor}.
As $n\rightarrow \infty ,$ $\forall x\in \left\{ 1,...,y-1\right\} $: 
\begin{equation}
\text{ }h\left( x\right) ={\bf E}h\left( X_{\tau _{x}}\right) =h\left(
0\right) {\bf P}\left( \tau _{x,0}<\tau _{x,y}\right) +{\bf E}h\left(
X_{\tau _{x,y}}\right) {\bf P}\left( \tau _{x,y}<\tau _{x,0}\right) .
\label{G17a}
\end{equation}
As $y\rightarrow \infty $, both $\tau _{x,y}$ and the overshoot $X_{\tau
_{x,y}}\rightarrow \infty $. Assuming there is a solution $h\left( x\right)
>0$ such that $h\left( x\right) \rightarrow 0$ as $x\rightarrow \infty $,
yields 
\begin{equation}
{\bf P}\left( \tau _{x,0}<\tau _{x,\infty }\right) =h\left( x\right) ,
\label{G18}
\end{equation}
indeed consistent with the guess: $h\left( x\right) >0$ and vanishing at $%
\infty $ when $x\rightarrow \infty $. This expression also shows that $%
h\left( x\right) $ must be decreasing with $x$. Note ${\bf P}\left( \tau
_{x,0}<\tau _{x,\infty }\right) ={\bf P}\left( X_{\tau _{x}}=0\right) $
where $\tau _{x}:=\tau _{x,0}\wedge \tau _{x,\infty }.$ We obtained:

\begin{proposition}
In the transient regime, with ${\bf h}$ defined in (\ref{G17}) obeying $%
h\left( 0\right) =1$, 
\[
h\left( x\right) ={\bf P}\left( \tau _{x,0}<\tau _{x,\infty }\right) ,
\]
is the probability of extinction starting from state $x$.
\end{proposition}

{\em Remark:} The function $\overline{{\bf h}}:={\bf 1}-{\bf h}$ clearly
also is a (increasing) harmonic function and so is any convex combination of 
${\bf h}$ and $\overline{{\bf h}}$ with weights summing to $1$, \cite{Neveu}%
. Clearly, $\overline{h}\left( x\right) ={\bf P}\left( \tau _{x,\infty
}<\tau _{x,0}\right) ={\bf P}\left( X_{\tau _{x}}=\infty \right) .$ $%
\triangleright $

\section{Related random walks}

We introduce and analyze two related random walks on the non-negative
integers.

\subsection{Life annuities policy}

Consider now the Markov chain with state-space ${\Bbb N}_{0}$ and with
temporal evolution{\bf :} 
\begin{equation}
X_{n+1}=X_{n}\wedge \delta _{n+1}+\beta _{n+1};\text{ }X_{0}>0,\text{ else}
\label{chain3}
\end{equation}
\[
\Delta X_{n}=X_{n+1}-X_{n}=-\left( X_{n}-\delta _{n+1}\right) _{+}+\beta
_{n+1}. 
\]
One of its notable observable is 
\[
Y_{n+1}=-\left( X_{n}-\delta _{n+1}\right) _{+}, 
\]
the amount of individuals that potentially could leave the system.

As before indeed, $X_{n}$ may represent the size of some population facing
steady demands (emigration) and supplies (immigration). But here, if the
demand exceeds $X_{n}$, there is no update of the population size whereas if
not, the population size is instantaneously switched to the lower size of
the demand. As a result, $X_{n}$ now represents the potential capacity of
response of a population (as the amount of individuals it would potentially
be able to release) when facing the demands and in the presence of additive
supplies. In this sense, (\ref{chain3}) is `dual' to (\ref{chain2}).

Alternatively, $X_{n}$ may represent life annuities indexed and ceilinged on
the stock price $\delta _{n+1}$ of some commodity balanced with an additive
automatic penalization itself indexed on the stock price $\beta _{n+1}$ of
some other commodity, in a stable and stationary environment. As before,
competing depletion and growth mechanisms can occur simultaneously.

In (\ref{chain3}), the i.i.d. sequence $\beta _{n}$ is again assumed to be
independent of $\delta _{n}$ (i.i.d.) and $X_{n-k}$ for all $k\geq 1.$ The
Markov chain (\ref{chain3}) was recently introduced in \cite{BS}, but
without the interpretations we give here of this model.\newline

{\em Remark:} (\ref{chain3}) suggests the continuous-time version of this
process 
\[
dX_{t}=-\left( X_{t_{-}}-\delta _{t}\right) _{+}\cdot dt+dN_{t} 
\]
where $\delta _{t}\stackrel{d}{=}\delta $ for all $t\geq 0$ is a stationary
process and $N_{t}$ a compound-Poisson process with jumps' amplitudes $\beta
\geq 0.$ $\triangleright $

The one-step stochastic transition matrix $P$ (obeying $P{\bf 1}={\bf 1}$,
where ${\bf 1}$ is a column vector of ones) of the Markov chain $\left\{
X_{n}\right\} $ in (\ref{chain3}) is: 
\begin{equation}
\begin{array}{l}
P\left( x,y\right) =\alpha \sum_{z=0}^{x-1}\overline{\alpha }^{z}b_{y-z}+%
\overline{\alpha }^{x}b_{y-x}\text{, }x\geq 1\text{ and }y\geq x \\ 
P\left( x,y\right) =\alpha \sum_{z=0}^{y}\overline{\alpha }^{z}b_{y-z}\text{%
, }x\geq 1\text{ and }y<x. \\ 
P\left( x,y\right) =\sum_{z=0}^{x}d_{z}b_{y-z}\text{, }x,y\geq 0.
\end{array}
\label{P3}
\end{equation}
In particular $P\left( x,0\right) =d_{0}b_{0}$ and $P\left( x,x\right)
=\alpha \sum_{z=0}^{x-1}\overline{\alpha }^{z}b_{x-z}+\overline{\alpha }%
^{x}b_{0}.$

For any given $X_{n}=x,$ $S\left( x\right) :=x\wedge \delta \leq x$ [with
again: $S\left( X_{n}\right) \leq X_{n}$ a.s.] has law now obtained as
follows: 
\begin{eqnarray*}
S\left( x\right) &\sim &\overline{\alpha }^{x}\varepsilon
_{x}+\sum_{y=0}^{x-1}\alpha \overline{\alpha }^{y}\varepsilon _{y} \\
{\bf E}\left( z^{S\left( x\right) }\right) &=&\overline{\alpha }%
^{x}z^{x}+\alpha \sum_{y=0}^{x-1}\overline{\alpha }^{y}z^{y}=\overline{%
\alpha }^{x}z^{x}+\alpha \frac{1-\left( \overline{\alpha }z\right) ^{x}}{1-%
\overline{\alpha }z}
\end{eqnarray*}
leading to the local drift and variance terms 
\begin{eqnarray*}
f\left( x\right) &:&={\bf E}\left( X_{n+1}-x\mid X_{n}=x\right) ={\bf E}%
\left( S\left( x\right) \right) -x+{\bf E}\left( \beta \right) =-x+\frac{%
\overline{\alpha }}{\alpha }\left( 1-\overline{\alpha }^{x}\right) +{\bf E}%
\left( \beta \right) \\
\sigma ^{2}\left( x\right) &:&=\sigma ^{2}\left( \left( X_{n+1}-x\right)
\mid X_{n}=x\right) =\frac{\overline{\alpha }}{\alpha }\left( 1-\overline{%
\alpha }^{x}\right) \left[ 1+\frac{\overline{\alpha }}{\alpha }\left( 1+%
\overline{\alpha }^{x}\right) \right] -2\frac{\overline{\alpha }}{\alpha }x%
\overline{\alpha }^{x}+\sigma ^{2}\left( \beta \right) .
\end{eqnarray*}
As $x$ is large, the walker `feels' a negative stabilizing linear drift $%
f\left( x\right) \sim -x$ whereas the variance $\sigma ^{2}\left( x\right)
\sim \frac{\overline{\alpha }}{\alpha ^{2}}+\sigma ^{2}\left( \beta \right)
=\sigma ^{2}\left( \delta \right) +\sigma ^{2}\left( \beta \right) $ goes to
a constant as $x\rightarrow \infty .$ Averaging over $X_{n}$, we get

\begin{equation}
\Phi _{n+1}\left( z\right) :={\bf E}\left( z^{X_{n+1}}\right) =\phi _{\beta
}\left( z\right) \sum_{x\geq 0}{\bf P}\left( X_{n}=x\right) \left( \overline{%
\alpha }^{x}z^{x}+\alpha \frac{1-\left( \overline{\alpha }z\right) ^{x}}{1-%
\overline{\alpha }z}\right)  \label{GG1}
\end{equation}
\begin{eqnarray*}
&=&\phi _{\beta }\left( z\right) \left( \Phi _{n}\left( \overline{\alpha }%
z\right) +\frac{\alpha }{1-\overline{\alpha }z}\left( 1-\Phi _{n}\left( 
\overline{\alpha }z\right) \right) \right) \\
&=&\phi _{\beta }\left( z\right) \frac{\alpha +\overline{\alpha }\left(
1-z\right) \Phi _{n}\left( \overline{\alpha }z\right) }{1-\overline{\alpha }z%
}
\end{eqnarray*}

Note $\Phi _{n+1}\left( 0\right) =b_{0}\left( \alpha +\overline{\alpha }\Phi
_{n}\left( 0\right) \right) $ entails 
\[
\Phi _{n}\left( 0\right) ={\bf P}\left( X_{n}=0\right) =b_{0}\alpha \frac{%
1-\left( b_{0}\overline{\alpha }\right) ^{n}}{1-b_{0}\overline{\alpha }}%
+\left( b_{0}\overline{\alpha }\right) ^{n}\Phi _{0}\left( 0\right) 
\]

\[
\stackunder{n\rightarrow \infty }{\rightarrow }{\bf P}\left( X_{\infty
}=0\right) =a:=\frac{b_{0}\alpha }{1-b_{0}\overline{\alpha }}, 
\]
a well-defined limit. The chain (\ref{chain3}) being irreducible and
aperiodic, it is sufficient for its positive-recurrence to show that $\Phi
_{n}\left( 0\right) $ approaches a positive limit as $n\rightarrow \infty $.
As a result, we get:

\begin{proposition}
The chain (\ref{chain3}) is positive recurrent (ergodic) with $\Phi _{\infty
}\left( 1\right) ={\bf P}\left( X_{\infty }<\infty \right) =1$, whatever the
law of $\beta $.
\end{proposition}

The chain (\ref{chain3}) is so strongly and rapidly attracted to $\left\{
0\right\} $ that there is no heavy-tailed r.v.'s capable to make it switch
to a transient regime, a remarkable stability property of this process. This
property may be viewed as a consequence of Proposition $4$ in \cite{BS},
stating that, with $\delta _{m}$ i.i.d. geometric$\left( \alpha \right) -$%
distributed, the iterate of the shrinking operator obeys 
\[
\wedge _{m=1}^{n}\left( \delta _{m+1}\wedge X_{m}\right) =D_{n+1}\wedge
_{m=1}^{n}X_{m} 
\]
where $D_{n+1}$ is a r.v. with geometric distribution having failure
probability $1-\overline{\alpha }^{n}$ tending to $1$ geometrically fast.

The limiting p.g.f. $\Phi _{\infty }\left( z\right) $ solves 
\begin{equation}
\Phi _{\infty }\left( z\right) =\phi _{\beta }\left( z\right) \frac{\alpha +%
\overline{\alpha }\left( 1-z\right) \Phi _{\infty }\left( \overline{\alpha }%
z\right) }{1-\overline{\alpha }z}  \label{Feq}
\end{equation}
If $\mu _{\beta }<\infty $, then $m:={\bf E}\left( X_{\infty }\right)
<\infty $ with $m=\mu _{\beta }+\frac{\overline{\alpha }}{\alpha }\left(
1-\Phi _{\infty }\left( \overline{\alpha }\right) \right) $

In one case and for a specific value of $b_{0}$, the p.g.f. $\Phi _{\infty
}\left( z\right) $ is geometric$\left( a\right) $ with the limiting failure
probability $a=\frac{b_{0}\alpha }{1-b_{0}\overline{\alpha }}\in \left(
0,1\right) $. This occurs when 
\[
\phi _{\beta }\left( z\right) =b_{0}+\overline{b}_{0}\frac{\alpha z}{1-%
\overline{\alpha }z}, 
\]
the p.g.f. of a zero-inflated geometric distribution, provided $%
b_{0}=1/\left( 1+\overline{\alpha }\right) \in \left( 0,1\right) $ in which
case $a=\alpha $. Indeed, 
\[
\phi _{\beta }\left( z\right) =\frac{1}{1+\overline{\alpha }}\frac{1-%
\overline{\alpha }^{2}z}{1-\overline{\alpha }z} 
\]
with mean $\mu _{\beta }=\overline{\alpha }/\left[ \alpha \left( 1+\overline{%
\alpha }\right) \right] =\overline{\alpha }/\left( 1-\overline{\alpha }%
^{2}\right) $ and $\Phi _{\infty }\left( z\right) =\frac{\alpha }{1-%
\overline{\alpha }z}$ solves (\ref{Feq}). $\Phi _{\infty }\left( z\right) $
is a one-parameter geometric solution with failure parameter $\alpha $ and
mean $\overline{\alpha }/\alpha $. We note that both $\phi _{\beta }\left(
z\right) $ and $\Phi _{\infty }\left( z\right) $ share the same algebraic
singularity at $z_{b}=1/\overline{\alpha }$. Except for this exceptional
case, the functional equation (\ref{Feq}) has no known explicit solution.
However, using singularity analysis, it is possible to extract a large $x$
expression of ${\bf P}\left( X_{\infty }=x\right) $ in the case of
algebraico-logarithmic singularity of $\phi _{\beta }\left( z\right) .$%
\newline

{\bf A short reminder on singularity analysis}, \cite{Fla}{\bf . }Let $\phi
\left( z\right) $ be any analytic function in the indented domain defined by

\[
D=\left\{ z:\left| z\right| \leq z_{0},\left| \text{Arg}\left(
z-z_{c}\right) \right| >\pi /2-\eta \right\} 
\]
where $z_{c}$, $z_{0}>z_{c}$, and $\eta $ are positive real numbers. Assume
that, with $\sigma \left( x\right) =x^{a}\left( \log x\right) ^{b}$ where $a$
and $b$ any real numbers (the singular exponents), we have

\begin{equation}
\phi \left( z\right) \sim C_{1}+C_{2}\sigma \left( \frac{1}{1-z/z_{c}}%
\right) \text{ as }z\rightarrow z_{c}\text{ in }D,  \label{for65}
\end{equation}
for some real constants $C_{1}$ and $C_{2}.$ Then:

- if $a\notin \left\{ 0,-1,-2,...\right\} $ the $z^{x}$-coefficients in the
power-series expansion of $\phi \left( z\right) $ satisfy

\begin{equation}
\left[ z^{x}\right] \phi \left( z\right) \sim C_{2}z_{c}^{-x}\cdot \frac{%
\sigma \left( x\right) }{x}\frac{1}{\Gamma \left( a\right) }\text{ as }%
x\rightarrow \infty \text{,}  \label{for66}
\end{equation}
where $\Gamma \left( a\right) $ is the Euler function. $\phi \left( z\right) 
$ presents an algebraic-logarithmic singularity at $z=z_{c}.$

- if $a\in \left\{ 0,-1,-2,...\right\} $ and $b\neq 0,$ the singularity $%
z=z_{c}$ is purely logarithmic and

\begin{equation}
\left[ z^{x}\right] \phi \left( z\right) \sim C_{2}bz_{c}^{-x}\cdot \frac{%
\sigma \left( x\right) }{x\log x}\left( \frac{1}{\Gamma }\right) ^{^{\prime
}}\left( a\right) \text{ as }x\rightarrow \infty \text{,}  \label{for67}
\end{equation}
involving the derivative of the reciprocal of the Euler gamma function at $a$%
.

Thus, for algebraic-logarithmic singularities, the asymptotic of the $z^{x}$%
-coefficients can be read from the singular behavior of the function $\phi
\left( z\right) $ under study.\newline

Coming back to (\ref{Feq}), we observe that if the p.g.f. $\phi _{\beta
}\left( z\right) $ has a singularity at some $z_{b}\geq 1$, it is also a
singularity for $\Phi _{\infty }\left( z\right) $. Indeed, the term $\frac{%
\alpha +\overline{\alpha }\left( 1-z\right) \Phi _{\infty }\left( \overline{%
\alpha }z\right) }{1-\overline{\alpha }z}\mid _{z=1/\overline{\alpha }}=1$
so that $1/\overline{\alpha }$ is not a singularity of $\Phi _{\infty
}\left( z\right) $ in (\ref{Feq}). The singularity analysis of $\phi _{\beta
}\left( z\right) $ can essentially be transfered to $\Phi _{\infty }\left(
z\right) $. \newline

{\bf Examples:}

- Suppose $a>0$, $\sigma \left( x\right) =x^{a}$ and 
\[
\phi _{\beta }\left( z\right) \sim C_{1}+C_{2}\left( 1-z/z_{c}\right) ^{-a}%
\text{ as }z\rightarrow z_{c}>1. 
\]
Then, as $z\rightarrow z_{c},$%
\begin{eqnarray*}
\Phi _{\infty }\left( z\right) &\sim &\left( C_{1}+C_{2}\left(
1-z/z_{c}\right) ^{-a}\right) \frac{\alpha +\overline{\alpha }\left(
1-z_{c}\right) \Phi _{\infty }\left( \overline{\alpha }z_{c}\right) }{\alpha
+\overline{\alpha }\left( 1-z_{c}\right) }\sim C_{1}+C_{2}^{\prime }\left(
1-z/z_{c}\right) ^{-a} \\
\left[ z^{x}\right] \Phi _{\infty }\left( z\right) &=&{\bf P}\left(
X_{\infty }=x\right) \sim C_{2}^{\prime }z_{c}^{-x}\cdot x^{a-1}\frac{1}{%
\Gamma \left( a\right) }\text{,}
\end{eqnarray*}
and $\Phi _{\infty }\left( z\right) $\ behaves similarly as $\phi _{\beta
}\left( z\right) $. The singularity of $\Phi _{\infty }\left( z\right) $ is
also the one $z_{c}>1$\ of $\phi _{\beta }\left( z\right) $\ because $\Phi
_{\infty }\left( \overline{\alpha }z_{c}\right) <1$. As a result, ${\bf P}%
\left( X_{\infty }=x\right) $ and ${\bf P}\left( \beta =x\right) $ have
similar behaviors for large $x$.

- If, with $\sigma \left( x\right) =x^{-a}$, $\phi _{\beta }\left( z\right)
=1-\lambda \left( 1-z\right) ^{a}$, $a,\lambda \in \left( 0,1\right) $,
where $z_{c}=1$. Then, as $z\rightarrow z_{c}=1,$ $\Phi _{\infty }\left(
z\right) \sim 1-\lambda \left( 1-z\right) ^{a}$ with $m=\infty .$ This is
because $\frac{\alpha +\overline{\alpha }\left( 1-z\right) \Phi _{\infty
}\left( \overline{\alpha }z\right) }{\alpha +\overline{\alpha }\left(
1-z\right) }\rightarrow 1$ as $z\rightarrow 1$. Here,

\[
\left[ z^{x}\right] \Phi _{\infty }\left( z\right) ={\bf P}\left( X_{\infty
}=x\right) \sim -\lambda x^{-\left( a+1\right) }\frac{1}{\Gamma \left(
-a\right) },\text{ as }x\rightarrow \infty , 
\]
with a power-law tail.

- Supposing, with $\sigma \left( x\right) =\left( \log x\right) ^{-b},$ $%
b>0, $%
\[
\phi _{\beta }\left( z\right) \stackunder{z\rightarrow zc}{\sim }C_{1}+\frac{%
C_{2}}{\left( -\log \left( 1-z/z_{c}\right) \right) ^{b}}, 
\]
then $\Phi _{\infty }\left( z\right) $\ behaves similarly as $\phi _{\beta
}\left( z\right) $ and $\left[ z^{x}\right] \Phi _{\infty }\left( z\right) =%
{\bf P}\left( X_{\infty }=x\right) \sim C_{2}^{\prime }bz_{c}^{-x}\cdot 
\frac{1}{x\left( \log x\right) ^{b+1}}\left( \frac{1}{\Gamma }\right)
^{^{\prime }}\left( 0\right) $; when $z_{c}=1$, both $\beta ,X_{\infty }$
have logarithmic moments of order smaller than $b$ but no moments of any
arbitrary positive order. $\Box $\newline

With $X_{0}=x_{0}\geq 1$, defining the double generating function 
\[
\Phi _{x_{0}}\left( u,z\right) =\sum_{n\geq 0}u^{n}\Phi _{n}\left( z\right)
=\sum_{n\geq 0}u^{n}{\bf E}_{x_{0}}\left( z^{X_{n}}\right) , 
\]
from (\ref{GG1}), we get

\begin{proposition}
\[
\Phi _{x_{0}}\left( u,z\right) =z^{x_{0}}+\frac{u\phi _{\beta }\left(
z\right) }{1-\overline{\alpha }z}\left[ \frac{\alpha }{1-u}+\overline{\alpha 
}\left( 1-z\right) \Phi _{x_{0}}\left( u,\overline{\alpha }z\right) \right] .
\]

Consequently, 
\[
\Phi _{x_{0}}\left( u,0\right) =G_{x_{0},0}\left( u\right) =\frac{%
b_{0}\alpha u}{\left( 1-u\right) \left( 1-b_{0}\overline{\alpha }u\right) }=%
\frac{b_{0}\alpha u}{1-b_{0}\overline{\alpha }}\left( \frac{1}{1-u}-\frac{%
b_{0}\overline{\alpha }}{1-b_{0}\overline{\alpha }u}\right) 
\]
is the Green kernel of this model and it is independent of $x_{0}.$
\end{proposition}

Indeed, whatever $x_{0}$, the only way to move to $0$ in one step has
probability $b_{0}\alpha $. The contact probability at $0$ is given by 
\[
\left[ u^{n}\right] \Phi _{x_{0}}\left( u,0\right) ={\bf P}_{x_{0}}\left(
X_{n}=0\right) =\frac{b_{0}\alpha }{1-b_{0}\overline{\alpha }}\left(
1-\left( b_{0}\overline{\alpha }\right) ^{n}\right) \stackunder{n\rightarrow
\infty }{\rightarrow }\frac{b_{0}\alpha }{1-b_{0}\overline{\alpha }}, 
\]
so the generic term of a non-summable series as required, translating that
state $0$ is visited infinitely often. The one-step probability of reaching $%
0$ being independent of the current state gives

\begin{corollary}
The first hitting time $\tau _{x_{0},0}$ of $0$ when starting from $%
x_{0}\geq 1$, is independent of $x_{0}$ and geometrically distributed with
failure probability $b_{0}\alpha $.
\end{corollary}

\subsection{Sub-critical BGW processes with immigration: massive depletion}

For comparison, we revisit these related bi-stochastic processes where a
similar competition between birth and death events holds. In such processes,
the shrinkage part of the population size is not due to the facing of
external demands of emigrants, rather by internal unbalance when on average
the branching number per capita is less than 1 (sub-criticality). This
depletion mechanism is much stronger than the one in (\ref{chain2}) and so
the conditions on $\beta $ under which (\ref{chain4}) is ergodic are much
weaker than the ones of the Section 2 model.

Bienyam\'{e}-Galton-Watson (BGW) processes with random branching number $\nu 
$ obeying ${\bf P}\left( \nu =0\right) >0$ are very unstable, going either
extinct or drifting to $\infty ,$ \cite{Harris}. Almost sure extinction of a
(sub-)critical BGW process can be avoided while allowing immigration. Let $%
\left( \beta _{n};n\geq 1\right) $ be an i.i.d. random sequence of r.v.'s
taking values in ${\Bbb N}_{0}$ (the immigrants) and let $\phi _{\beta
}\left( z\right) ={\bf E}\left( z^{\beta }\right) $ be their common p.g.f..
With $X_{0}=0$, consider the Markov branching process with immigration
(BGWI) $X_{n}$, recursively defined by 
\begin{equation}
X_{n+1}=\sum_{i=1}^{X_{n}}\nu _{i}+\beta _{n+1}\text{, }n\geq 0.
\label{chain4}
\end{equation}
Here the $\nu _{i}$'s are the i.i.d. branching numbers of the underlying BGW
process. We let $\phi _{\nu }\left( z\right) :={\bf E}\left( z^{\nu }\right) 
$. With $\left[ z^{y^{\prime }}\right] \phi _{\nu }\left( z\right) $ the $%
z^{y^{\prime }}$-coefficient of $\phi _{\beta }\left( z\right) $, $\left\{
X_{n}\right\} $ has one-step transition matrix ($x,y\in {\Bbb N}_{0}$): 
\[
P\left( x,y\right) =\sum_{y^{\prime }=0}^{y}\left[ z^{y^{\prime }}\right]
\phi _{\nu }\left( z\right) ^{x}b_{y-y^{\prime }}. 
\]

{\em Remark:} Associated to (\ref{chain4}) is the continuous-time version of
this Markov process 
\[
dX_{t}=\sum_{i=1}^{X_{t_{-}}}\left( \nu _{i}-1\right) \cdot dt+dN_{t}, 
\]
where $N_{t}$ a compound-Poisson process with jumps' amplitudes $\beta .$ $%
\triangleright $

We assume the underlying BGW is subcritical viz. $\mu _{\nu }:=\phi _{\nu
}^{\prime }\left( 1\right) ={\bf E}\left( \nu \right) <1$. With $S\left(
x\right) :=\sum_{i=1}^{x}\nu _{i}$ the shrinking part of the dynamics (\ref
{chain4}), given $X_{n}=x$, we now have the local drift and variance terms 
\begin{eqnarray*}
f\left( x\right) &:&={\bf E}\left( X_{n+1}-x\mid X_{n}=x\right) ={\bf E}%
\left( S\left( x\right) \right) -x+{\bf E}\left( \beta \right) =-x\left(
1-\mu _{\nu }\right) +{\bf E}\left( \beta \right) \\
\sigma ^{2}\left( x\right) &:&=\sigma ^{2}\left( \left( X_{n+1}-x\right)
\mid X_{n}=x\right) =\sigma ^{2}\left( S\left( x\right) \right) +\sigma
^{2}\left( \beta \right) =x\sigma ^{2}\left( \nu \right) +\sigma ^{2}\left(
\beta \right) \\
&\sim &x\sigma ^{2}\left( \nu \right) \text{ as }x\rightarrow \infty \text{
if }\sigma ^{2}\left( \nu \right) <\infty .
\end{eqnarray*}
As $x$ is large, the walker `feels' the negative drift $f\left( x\right) 
\stackunder{x\rightarrow \infty }{\sim }-x\left( 1-\mu _{\nu }\right) .$
Fluctuations are very large for large $x$, of order $x$, so of diffusive
type.

If the underlying BGW is subcritical, the adjunction of immigrants will
prevent its extinction in that the BGWI will in general stabilize to a
well-defined random limit. The condition is that \cite{Heath} 
\[
\int_{0}^{1}\frac{1-\phi _{\beta }\left( z\right) }{\phi _{\nu }\left(
z\right) -z}dz<\infty \text{,} 
\]
which is met if ${\bf E}\left( \nu \right) <1$ and ${\bf E}\log \left(
1+\beta \right) <\infty $ (a very weak existence condition of the
log-moments of $\beta $ invalidated only would $\beta $ have extremely heavy
logarithmic tails). Letting $\Phi _{n}\left( z\right) ={\bf E}\left(
z^{X_{n}}\right) ,$ 
\begin{eqnarray*}
\Phi _{n+1}\left( z\right) &=&\phi _{\beta }\left( z\right) \Phi _{n}\left(
\phi _{\nu }\left( z\right) \right) \text{, }\Phi _{0}\left( z\right) =1 \\
\Phi _{n}\left( z\right) &=&\prod_{m=0}^{n-1}\phi _{\beta }\left( \phi _{\nu
}^{\circ m}\left( z\right) \right)
\end{eqnarray*}
where $\phi ^{\circ m}\left( z\right) $ is the $m$-th iterate of $\phi _{\nu
}$ and $\Phi _{\infty }\left( z\right) $ obeys the functional equation $\Phi
_{\infty }\left( z\right) =\phi _{\beta }\left( z\right) \Phi _{\infty
}\left( \phi _{\nu }\left( z\right) \right) .$ Hence, under the above
condition, $\Phi _{\infty }\left( z\right) =\prod_{m=0}^{\infty }\phi
_{\beta }\left( \phi _{\nu }^{\circ m}\left( z\right) \right) $ is the
well-defined (convergent) p.g.f. of $X_{\infty }$, translating that $%
X_{\infty }$ is an infinite sum of independent r.v.'s with p.g.f.'s $\phi
_{\beta }\left( \phi _{\nu }^{\circ m}\left( z\right) \right) $, $m\geq 0$,
with mean ${\bf E}\left( X_{\infty }\right) =\phi _{\beta }^{\prime }\left(
1\right) /\left( 1-\phi _{\nu }^{\prime }\left( 1\right) \right) \leq \infty
.$ Note $X_{\infty }\stackrel{d}{=}\beta +E$ where $E$ is an excess r.v.
independent of $\beta .$ Equivalently, $X_{\infty }$ is characterized by ($%
X_{\infty }^{\prime }$ denoting a statistical copy of $X_{\infty }$) 
\[
X_{\infty }\stackrel{d}{=}\sum_{i=1}^{X_{\infty }^{\prime }}\nu _{i}+\beta , 
\]
with $\beta $ independent of $\sum_{i=1}^{X_{\infty }^{\prime }}\nu _{i}$,
translating its generalized self-decomposability property induced by the
semi-group generated by $\phi _{\nu }$ (see \cite{AB} and \cite{SH}, Section
V.8).\newline

Some famous examples are:

- The pure-death case: this is when $\nu \stackrel{d}{\sim }$Bernoulli$%
\left( c\right) ,$ equivalently $\phi _{\nu }\left( z\right) =1-c+cz,$ with $%
c\in \left( 0,1\right) $ the survival probability. In that case (alluded to
in the Introduction), $S\left( x\right) \stackrel{d}{\sim }$bin$\left(
x,c\right) $. This binomial shrinkage effect is appropriate when the
individuals of the current population each die or survive{\bf \ }in an
independent way with some survival probability $c$. There is thus a
resulting drastic depletion of individuals at each step. The BGW process
with this branching mechanism is necessarily subcritical. In that case 
\[
\phi _{\nu }^{\circ m}\left( z\right) =1-c^{m}+c^{m}z 
\]
is itself Bernoulli with survival probability $c^{m}.$ Assuming a
discrete-stable$\left( \alpha ,\theta \right) $ distribution for $\beta $, $%
\alpha \in \left( 0,1\right] $, $\theta >0$, [with scale parameter $\theta $
and power-law tails of index $\alpha $] it holds that $\Phi _{\infty }\left(
z\right) =e^{-\theta /\left( 1-c^{\alpha }\right) \left( 1-z\right) ^{\alpha
}}$, so $X_{\infty }$ is itself $\alpha -$discrete-stable$\left( \alpha
,\theta /\left( 1-c^{\alpha }\right) \right) $, but with scale parameter $%
\theta /\left( 1-c^{\alpha }\right) .$ We have ${\bf E}\left( X_{\infty
}\right) =\infty $ if $\alpha \in \left( 0,1\right) $ and $\alpha =1$ is the
Poisson$\left( \theta \right) $ special case with ${\bf E}\left( X_{\infty
}\right) =\theta /\left( 1-c\right) <\infty $.

- The simple birth and death case (binary fission) is when $\phi _{\nu
}\left( z\right) =\pi _{0}+\pi _{1}z+\pi _{2}z^{2}$, $\pi _{0}+\pi _{1}+\pi
_{2}=1$ (subcritical when $\pi _{2}<\pi _{0}$)$.$

- The linear-fractional case is when $\phi _{\nu }\left( z\right) =\pi _{0}+%
\overline{\pi }_{0}\frac{\pi z}{1-\overline{\pi }z}$, $\left( \pi _{0},\pi
\right) \in \left( 0,1\right) $, $\overline{\pi }_{0}=1-\pi _{0},$ $%
\overline{\pi }=1-\pi $ (subcritical when $\overline{\pi }_{0}/\pi <1$).%
\newline

In contrast with the truncated geometric model of Section $2$, the
ergodicity property of BGWI processes can be extended to an underlying {\em %
critical} BGW (with $\mu _{\nu }=1$) if \cite{Seneta} 
\[
{\bf E}\left( \beta \right) <\infty \text{ and }\int_{0}^{1}\frac{1-z}{\phi
_{\nu }\left( z\right) -z}dz<\infty , 
\]
translating that $\nu $ has an infinite variance, viz. $\phi _{\nu }\left(
z\right) \stackunder{z\rightarrow 1}{\sim }z+\frac{\lambda }{\alpha +1}%
\left( 1-z\right) ^{\alpha +1}+o\left( \left( 1-z\right) ^{\alpha +1}\right) 
$, with both $\lambda ,\alpha \in \left( 0,1\right) $. In that case, the
limit probability law of $X_{\infty }$ exists and it is heavy-tailed with
index $\alpha $ (so in particular with ${\bf E}\left( X_{\infty }\right)
=\infty $). If the infinite variance property of the underlying BGW process
is not satisfied, the BGWI process is transient and drifts to $\infty .$
Still, a non-finite invariant measure is known to exist in this case, \cite
{Seneta3}. \newline

In the supercritical regime ($\mu _{\nu }>1$) and if $\mu _{\beta }<\infty ,$
the BGWI process is transient at $\infty $ and it can be checked that, as $%
n\rightarrow \infty ,$%
\begin{eqnarray*}
\frac{1}{n}X_{n} &\rightarrow &0\text{ a.s. if }1<\mu _{\nu }<\infty \\
\frac{1}{n}X_{n} &\rightarrow &\infty \text{ a.s. if }\mu _{\nu }=\infty .
\end{eqnarray*}
If $1<\mu _{\nu }<\infty $, the sequence $\mu _{\nu }^{-n}X_{n}$ is a
converging bounded in $L^{1}$ nonnegative submartingale and so 
\[
\mu _{\nu }^{-n}X_{n}\rightarrow Z<\infty \text{ a.s.} 
\]
where the law of $Z$ is characterized by its Laplace-Stieltjes transform in
the main Theorem of \cite{Seneta2}. Invariant measures are known to exist in
the supercritical as well but they are non-unique, \cite{Seneta3}. The idea
is to come down to the non-supercritical case while allowing defective
immigration laws.

\section{Concluding remarks}

We first analyzed the recurrence/transience conditions of a Markov chain on
the nonnegative integers where systematic random immigration events
promoting growth were simultaneously balanced with random emigration based
on a truncated geometric rule, favoring shrinkage. Two related random walks
in the same spirit, but with different collapse rules, were introduced. One
is a `dual' version of the latter process, shown to be always stable
(recurrent), the other the classical BGW process with immigration where the
origin of the pruning is internal, rather due to sub-criticality of the
population. Their intrinsic statistical properties were shown to be of a
completely different nature: the first truncated geometric model turns out
to exhibit a large range of transience (the supercritical regime $\mu
_{\beta }>\mu _{\delta }$), the second dual one is never transient while a
sub-critical BGW process with immigration fails to be recurrent only when
the immigrants' law has infinite logarithmic moments, a situation which can
be considered as exceptional in concrete population model applications.{\bf %
\ }\newline

{\bf Declarations of interest}

The author has no conflicts of interest associated with this paper.

{\bf Data availability statement}

There are no data associated with this paper.\newline

{\bf Acknowledgments:} T.H. acknowledges partial support from the labex
MME-DII ({\it Mod\`{e}les Math\'{e}matiques et \'{E}conomiques de la
Dynamique, de l' Incertitude et des Interactions}), ANR11-LBX-0023-01. This
work also benefited from the support of the Chair ``{\it Mod\'{e}lisation
Math\'{e}matique et Biodiversit\'{e}}'' of Veolia-Ecole
Polytechnique-MNHN-Fondation X. This work was also funded by CY Initiative
of Excellence (grant ``{\it Investissements d'Avenir}''ANR- 16-IDEX-0008),
Project ``EcoDep'' PSI-AAP2020-0000000013.

The author is most grateful to his Referees and the Editor-in-Charge for
their suggestions to improve the presentation of this manuscript and most of
all for pointing out some computational errors in an early version of it.

\end{document}